\documentclass[authorversion,sigconf,screen]{acmart}

\setcopyright{licensedusgovmixed}
\acmPrice{15.00}
\acmDOI{10.1145/3503222.3507779}
\acmYear{2022}
\copyrightyear{2022}
\acmSubmissionID{asplos22main-p1787-p}
\acmISBN{978-1-4503-9205-1/22/02}
\acmConference[ASPLOS '22]{Proceedings of the 27th ACM International Conference on Architectural Support for Programming Languages and Operating Systems}{February 28 -- March 4, 2022}{Lausanne, Switzerland}
\acmBooktitle{Proceedings of the 27th ACM International Conference on Architectural Support for Programming Languages and Operating Systems (ASPLOS '22), February 28 -- March 4, 2022, Lausanne, Switzerland}

\usepackage[normalem]{ulem}

\usepackage{balance}
\usepackage{microtype}
\usepackage{appendix}
\usepackage{tikz}
\usepackage{amsmath}
\usepackage{enumitem}
\usepackage{paralist}
\usepackage{subcaption}
\usepackage{placeins}
\usepackage{booktabs}
\usepackage{hyphenat}
\usepackage{verbatimbox}

\graphicspath{{./pictures/}}
\graphicspath{{./charts/}}

\begin{document}

\title[Adelie: Continuous Address Space Layout Re-randomization for Linux Drivers]{Adelie: Continuous Address Space Layout Re-randomization for Linux Drivers}
\titlenote{The U.S. Government is authorized to reproduce and distribute reprints for Governmental purposes notwithstanding any copyright annotation thereon.}

\author{Ruslan Nikolaev}
\authornote{Most of the work was done while the author worked at Virginia Tech.}
\email{rnikola@psu.edu}
\affiliation{%
        \institution{The Pennsylvania State University}
        \city{University Park}
        \state{PA}
        \country{USA}
}

\author{Hassan Nadeem}
\email{hnadeem@vt.edu}
\affiliation{%
        \institution{Virginia Tech}
        \city{Blacksburg}
        \state{VA}
        \country{USA}
}

\author{Cathlyn Stone}
\email{stonecat@vt.edu}
\affiliation{%
        \institution{Virginia Tech}
        \city{Blacksburg}
        \state{VA}
        \country{USA}
}

\author{Binoy Ravindran}
\email{binoy@vt.edu}
\affiliation{%
        \institution{Virginia Tech}
        \city{Blacksburg}
        \state{VA}
        \country{USA}
}

\begin{abstract}

While address space layout randomization (ASLR) has been extensively studied for user-space programs, the corresponding OS kernel's KASLR support remains very limited, making the kernel vulnerable to just-in-time (JIT) return-oriented programming (ROP) attacks. Furthermore, commodity OSs such as Linux restrict their KASLR range to 32 bits due to architectural constraints (e.g., x86-64 only supports 32-bit immediate operands for most instructions), which makes them vulnerable to even unsophisticated brute-force ROP attacks due to low entropy. Most in-kernel pointers remain static, exacerbating the problem when pointers are leaked.

Adelie, our kernel defense mechanism, overcomes KASLR limitations, increases KASLR entropy, and makes successful ROP attacks on the Linux kernel much harder to achieve. First, Adelie enables the \emph{position-indepe\-ndent code} (PIC) model so that the kernel and its modules can be placed anywhere in the 64-bit virtual address space, at any distance apart from each other. Second, Adelie implements \emph{stack re-randomization} and \emph{address encryption} on modules. Finally, Adelie enables \emph{efficient continuous KASLR} for modules by using the PIC model to make it (almost) impossible to inject ROP gadgets through these modules regardless of gadget's origin.

Since device drivers (typically compiled as modules) are often developed by third parties and are typically less tested than core OS parts, they are also often more vulnerable. By fully re-randomizing device drivers, the last two contributions together prevent most JIT ROP attacks since vulnerable modules are very likely to be a starting point of an attack. Furthermore, some OS instances in virtualized environments are specifically designated to run device drivers, where drivers are the primary target of JIT ROP attacks. Using a GCC plugin that we developed, we automatically modify different kinds of kernel modules. Since the prior art tackles only user-space programs, we solve many challenges unique to the kernel code. Our evaluation shows high efficiency of Adelie's approach: the overhead of the PIC model is completely negligible and re-randomization cost remains reasonable for typical use cases.

\end{abstract}



\maketitle

\newcommand{\architecture}{
\begin{figure}[ht!]
\centering
\subfloat[Zero-copying mechanism]{
\includegraphics[width=.48\columnwidth]{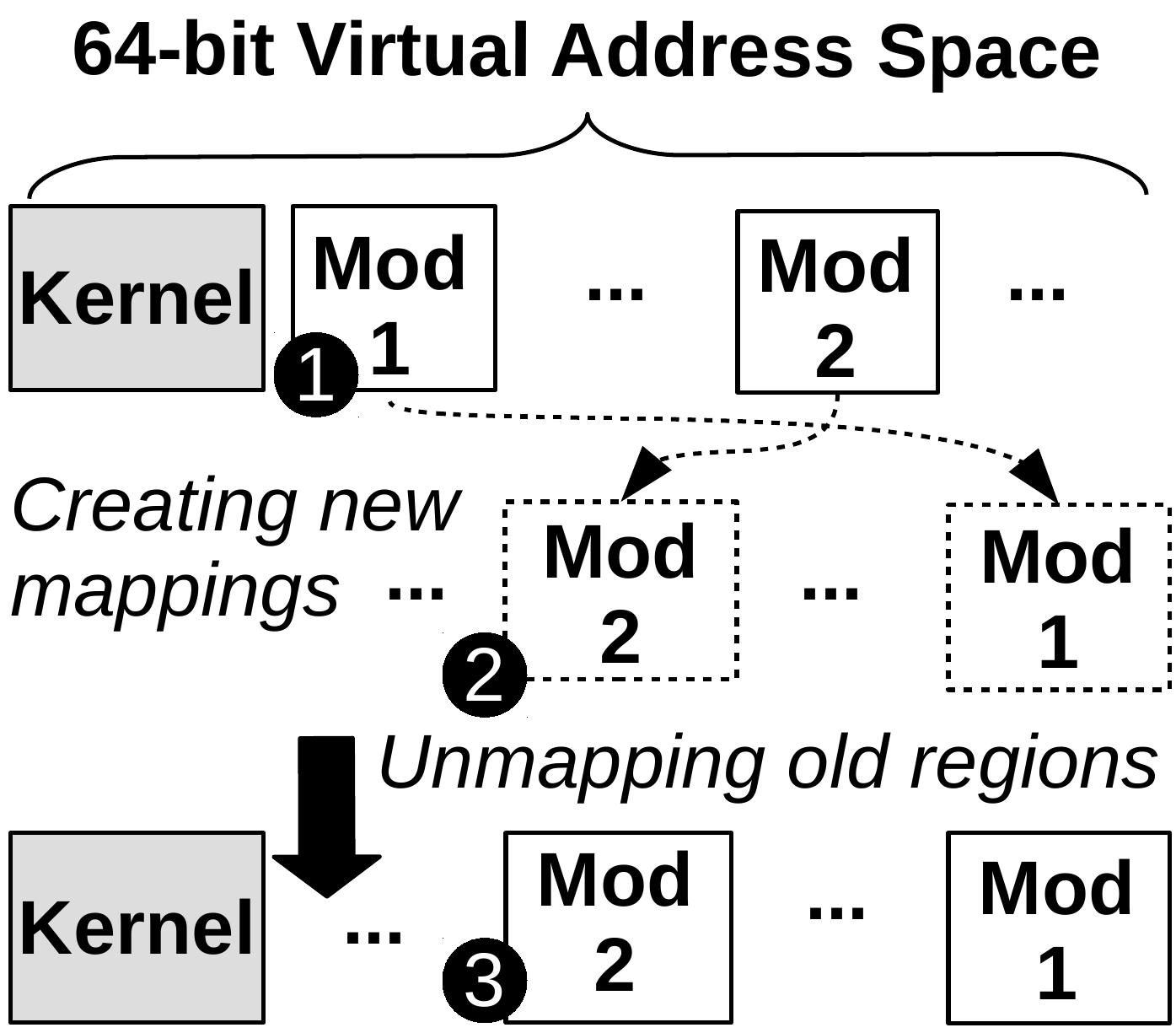}
\label{fig:architecture}
}
\subfloat[Re-randomizable modules]{
\includegraphics[width=.48\columnwidth]{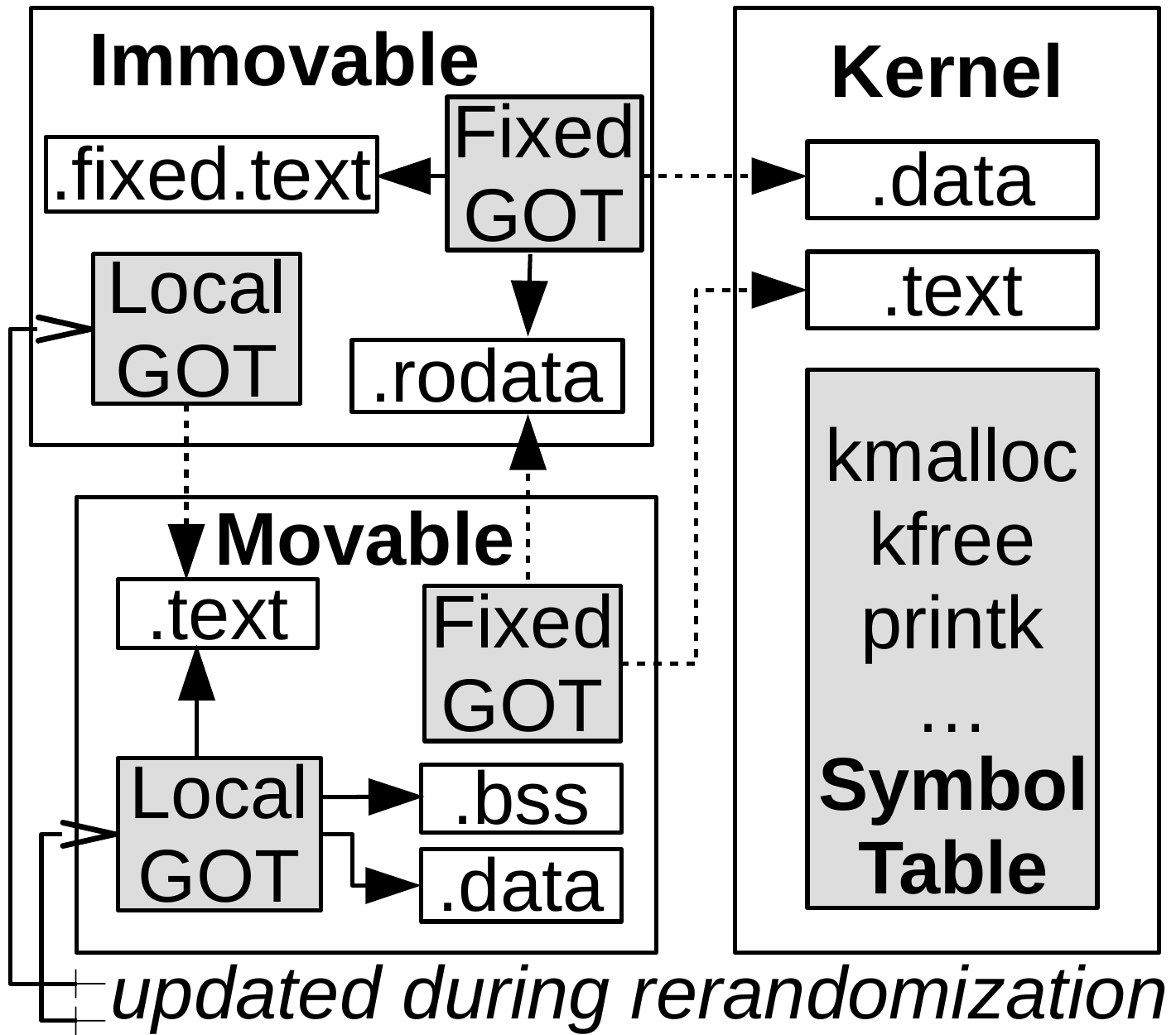}
\label{fig:design}
}
\caption{Adelie's design.}
\end{figure}
}

\newcommand{\wrapper}{
\begin{figure}[ht!]
\centering
\subfloat[Externally accessible functions]{
\includegraphics[width=.5\columnwidth]{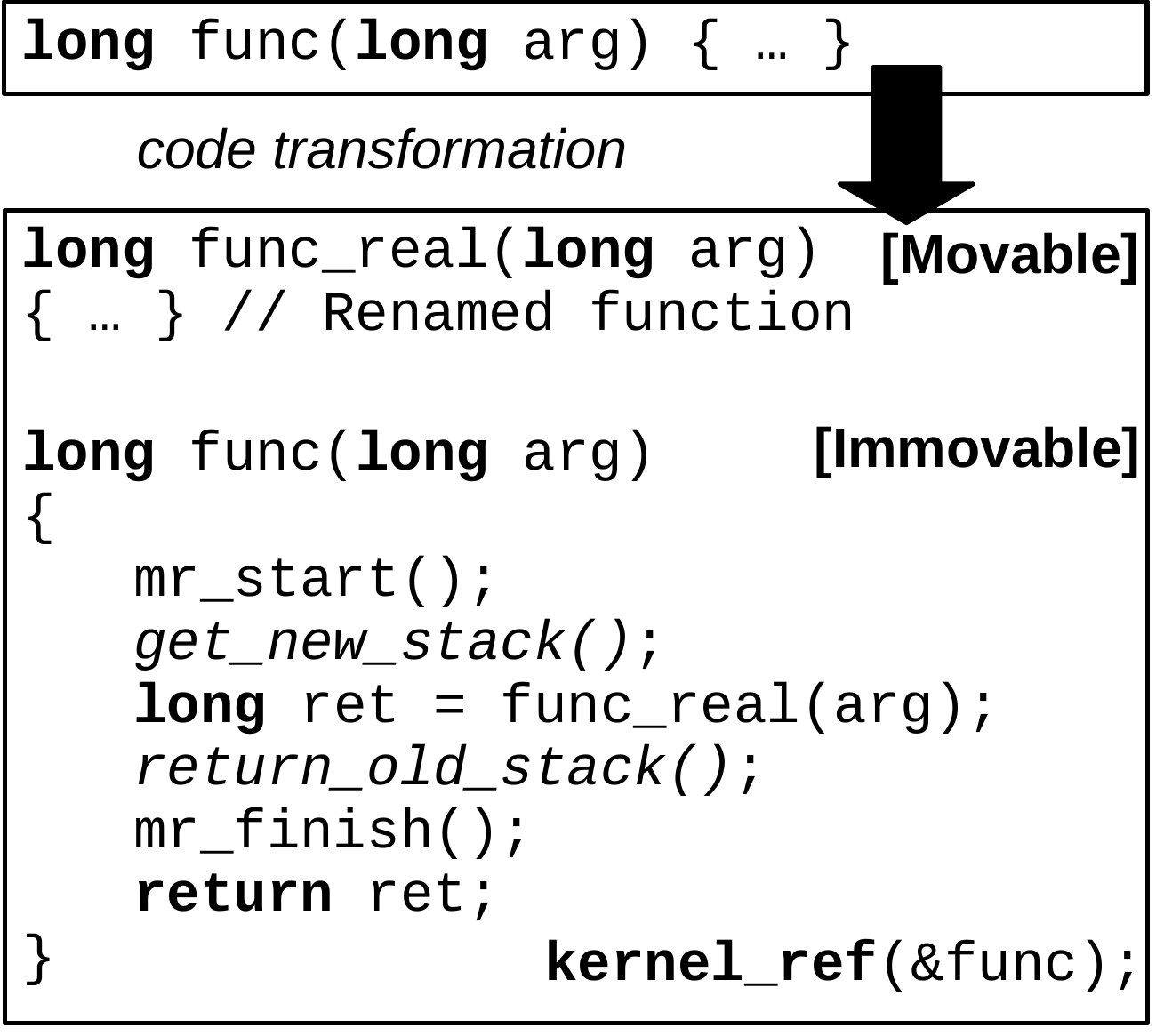}
\label{fig:wrapper}
}
\subfloat[Stack re-randomization]{
\includegraphics[width=.465\columnwidth]{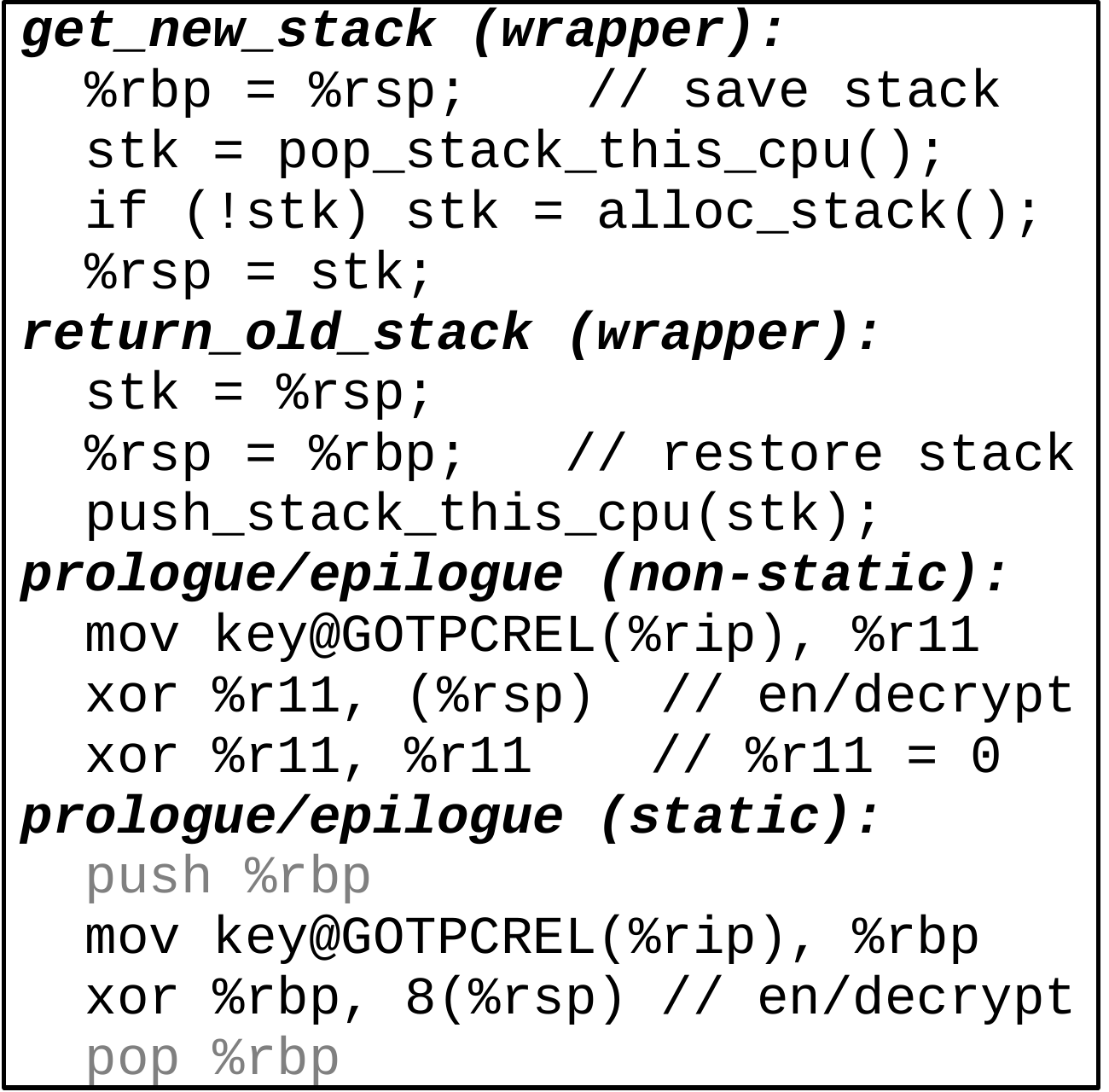}
\label{fig:stack}
}
\caption{Code transformations.}
\end{figure}
}

\section{Introduction}

\begin{figure}
        \centering
        \includegraphics[width=.8\columnwidth]{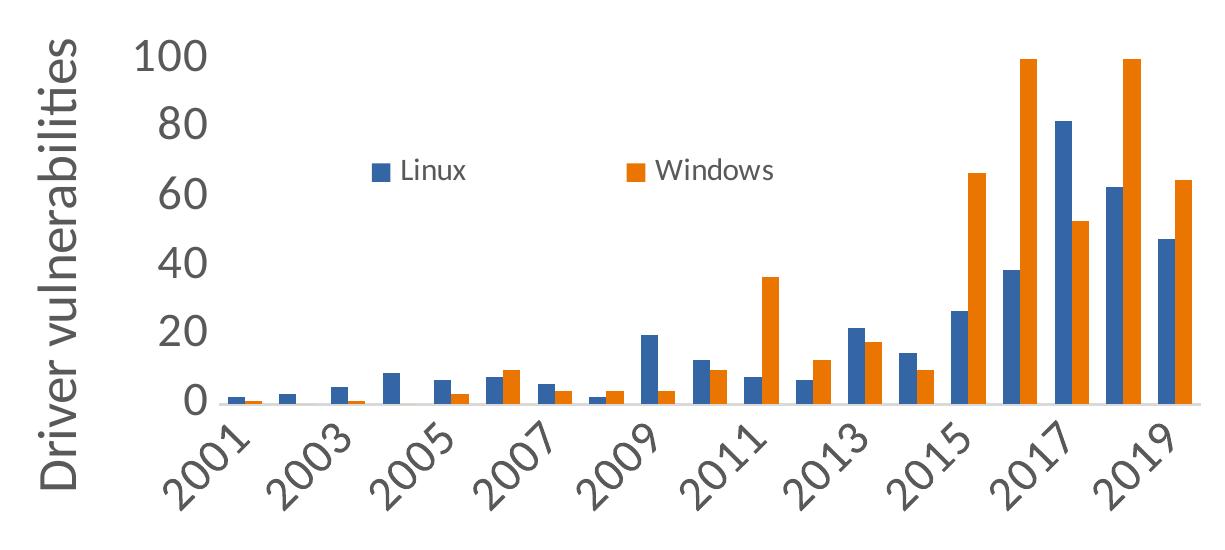}
			\caption{CVEs for device drivers~\cite{CVES}.}
   \label{fig:cve_driver}
    \end{figure}

As the sophistication of security attacks and countermeasures grows for user
space programs, OS kernels attract an increasing number of attackers, with an ever increasing number of kernel vulnerabilities
being discovered in the code of popular commodity OSs~\cite{LINUXVULN,MACOSVULN,WINDOWSVULN}. Certain vulnerabilities such as CVE-2018-14634 are very serious and seem to have existed for over a decade~\cite{LINUXVULNDECADE}.
Vulnerabilities -- the starting points of attacks -- are even more likely to be present in device drivers~\cite{10.1145/2103799.2103805,10.1145/502059.502042,10.1145/3339252.3340506,10.1145/1961296.1950401} since they are typically not as rigorously tested as core kernel components, as each installation uses only a subset of drivers. Moreover, the number of common vulnerabilities and exposures (CVE) calculated specifically for drivers continues to increase across different OSs exponentially (Figure~\ref{fig:cve_driver}).

There are several reasons why OS kernels are attractive to attackers. First and foremost, defense against attacks is typically more challenging in the kernel space, which includes low-level code involving system calls, device drivers, and interrupt handlers. Second, defense mechanisms for  kernel space vulnerabilities (and their inclusion in commodity OSs) often lag behind their user space counterparts. A case in point: address space layout randomization (ASLR) for the Linux kernel~\cite{KASLRLINUX},
a well-known technique against control-flow attacks, is limited to a paltry $2$GB range on x86-64 due to instruction immediate operand constraints~\cite{X86ABI}. Attackers can further assume that certain addresses are page-aligned, which makes even simple brute-force attacks feasible.\footnote{For 4KB pages, an attacker needs $\le 2^{31-12}=512$K attempts, likely $<1$K if the attacker knows an OS version, etc. An early boot-time (brute-force) attack, when a FS journal (ext4) is not yet flushed, will leave no traces at all.}

Furthermore, OS kernel code, especially that of typical OSs in widespread production-use such as Linux, is large and complex. On the one hand, this complicates the design and implementation of defense mechanisms~\cite{ASRMINIX}. One the other hand, it gives great flexibility for an attacker. In fact, there is a great reward
for an attacker: tampering with the kernel and gaining control of the entire system effectively enables the attacker to bypass many defense mechanisms that are deployed in user space to protect applications.

We present Adelie, an OS kernel defense mechanism that contributes to Linux security in several uniquely distinguishing ways. First, it extends kernel ASLR (KASLR) for the entire 64-bit address space efficiently by using position-independent code (PIC).
Second, it implements stack re\hyp{}randomization and address encryption
techniques.
Finally, it implements continuous address space re\hyp{}randomization.

Adelie's first contribution directly enhances the security of the Linux kernel
and its modules and helps to efficiently implement techniques in the other contributions.
Since drivers are more likely to have vulnerabilities than core kernel code~\cite{10.1145/2103799.2103805,10.1145/502059.502042,10.1145/3339252.3340506,10.1145/1961296.1950401} and because modules expose most ROP gadgets (see Section~\ref{sec:analysis}), we scope our last two contributions
to kernel modules only and the most vulnerable of them -- device drivers.
As we further discuss in Section~\ref{sec:analysis}, 
this targeted practical approach which incurs very little overhead suffices as
all these techniques combined (continuous ASLR, stack re\hyp{}randomization, and encryption)
prevent ROP attacks even in the presence of gadgets which originate in the core kernel or non-rerandomized modules.
Thus, Adelie's second and third contributions together provide a strong defense against just-in-time (JIT) ROP attacks~\cite{JITROP} in the entire Linux kernel.

Adelie's mechanisms are designed specifically for kernel space and 
solve unique challenges which are not present in prior art for user space~\cite{CODEARMOR,SHUFFLER}.
Adelie uses a zero-copying method for moving code and static data. In addition, Adelie efficiently keeps track of and unmaps previously used addresses.

We evaluate Adelie's PIC support using Sysbench, Kernbench, and microbenchmarks that characterize OS-heavy workloads.
We also evaluate the cost of re\hyp{}randomization for several drivers
using microbenchmarks and real-life server applications (Apache and mySQL). Adelie's re-randomization overhead is small ($<$ 2\%) and entirely negligible for supporting 64-bit KASLR with PIC modules.

\textbf{The paper makes the following contributions:}
\begin{compactenum}[1.]
\item Compiling and running all modules from the Linux kernel tree
(e.g., over 5000 modules in Ubuntu 18.04 which we used for testing) as
	PIC to extend kernel's ASLR range.\footnote{This contribution was submitted to one of the Linux kernel mailing lists as a series of patches: see \url{https://www.openwall.com/lists/kernel-hardening/2019/03/21/6} and \url{https://www.openwall.com/lists/kernel-hardening/2019/03/21/7}.}

\item Mechanisms for stack re\hyp{}randomization, address encryption, and continuous ASLR on Linux modules.
Our work is the first to target
large-scale OS kernels such as Linux.
Due to the gargantuan engineering effort required, plus legacy and low-level code in the kernel, we ruled out re\hyp{}randomization
for the entire kernel. Re-randomization
also incurs extra overheads. For these practical reasons, we re\hyp{}randomize only on the most vulnerable components. We argue that such a targeted approach is not only justified from a performance perspective, but is also more likely to be upstreamed into mainline Linux.
Vulnerable drivers are a likely starting point of an attack.
As further elaborated in Section~\ref{sec:analysis}, our approach
achieves strong protection against ROP gadget injection for the entire kernel
ecosystem regardless
of whether gadgets (e.g., in the core kernel which is not re-randomized) still
exist.
We implement a GCC plugin~\cite{gccPlugins}, which automatically converts existing modules to re-randomizable modules.

\item Demonstrating the mechanisms and their generality using
notable drivers compiled as re-randomizable modules.

\end{compactenum}

\textbf{Adelie is deployed in a real-life system. }
When using virtualization and dividing the system
into multiple guest OSs, device drivers can run either in a privileged virtual machine (VM), known as Dom0 in Xen, or in dedicated VMs, known as driver VMs
in Xen. When using driver VMs, drivers
are fundamentally the \emph{only} vulnerable components in the corresponding
guest OS. We used Adelie in an open-source enterprise system~\cite{SAVIOR} to
re-randomize a network driver in Dom0. This system comprises multiple VMs.

\section{Background}

In this section, we discuss control-flow attacks, remedies against them, and the specifics of ELF binaries and PIC.

\subsection{Return-Oriented Programming (ROP)}

Typical attacks modify the control-flow state of the program by executing an
unintended sequence of instructions. Since data-bound checks are not always
enforced by a programmer, an attacker can exploit
this vulnerability using buffer overflow. Such attacks generally overwrite function return addresses on the stack, thereby hijacking the program's control flow.
Modern CPUs support the Write-XOR-Execute feature, known as the NX
(Non-Execute) bit in x86-64~\cite{NXLINUX}, to prevent a memory page from being both
writable and executable. Data pages are marked as NX nowadays~\cite{UBUNTUNX}, making direct code injection impossible.

However, the NX mitigation can be bypassed via code reuse attacks such as
\textit{return-oriented programming
  (ROP)}~\cite{Bletsch:2011:JPN:1966913.1966919,Checkoway:2010:RPW:1866307.1866370,Roemer:2012:RPS:2133375.2133377,Shacham:2007:GIF:1315245.1315313}.
In these attacks, a stack buffer overflow is exploited to overwrite
a return address on the stack with the address of a selected function (e.g., from \textit{libc}). 
By carefully overwriting the stack, an attacker chains the
execution of a set of functions.
A variation of this technique, \textit{stack pivoting}, manipulates the
stack pointer register to point to memory where the crafted 
payload resides~\cite{Prakash:2015:DRT:2818000.2818023}.
Given a large enough set of loaded instructions,
it is possible to identify a control flow instruction, such as a return
or indirect jump, and create a sequence of
valid instructions (\textit{``gadget''}), which, when executed, yields
a desired behavior.
Jump-Oriented Programming (JOP) is a
variant of ROP, where a jump instruction alters control
flow~\cite{Bletsch:2011:JPN:1966913.1966919,Checkoway:2010:RPW:1866307.1866370}.

ROP's fundamental premise is that given a large enough set of already loaded
instructions or even arbitrary executable bytes, one
can piece together a sequence of instructions that is functionally
valid on a given ISA and accomplishes the desired goal. This is because of the density of ISAs, in
particular ISAs such as x86-64. 
By exploiting a buffer overflow, an attacker can therefore overwrite a
stack return address with the address of the first instruction of a
gadget, whose last instruction in turn overwrites the stack return
address with the first instruction of another gadget, thereby
executing a chain of gadgets which yields arbitrary program behavior from
existing code.

ROP is Turing-complete~\cite{Buchanan:2008:GIG:1455770.1455776} --
i.e., given a sufficiently large binary, any
functionality can be emulated by chaining
gadgets. Several tools have been developed to
automatically create ROP payloads from program
binaries~\cite{ropgadget,ROPCOMPILER} -- e.g., the
ROPgadget tool~\cite{ropgadget} can create an attack payload that
spawns a shell that can accept arbitrary commands from an attacker.

Even though attackers rarely use ROP alone, preferring to use
it only as a bridge to accessing more direct means of controlling
execution, existence of such tools and ROP compilers shows that 
the opportunity for arbitrarily powerful unintended execution is 
not rare but inherent in binary code, and must be mitigated.

A related feature, SMAP (Supervisor Mode Access Prevention), available
on recent x86-64 CPUs and supported by Linux, prevents the OS
kernel from being tricked to use data or code from user space. Adelie assumes
this feature is enabled.

\subsection{Protection against Control-Flow Attacks}

Code randomization is a common technique to defend against
control-flow attacks. ASLR~\cite{PAXASLR,ASLR} 
is a well-known technique used in modern OSs to protect user-space programs
by randomizing the memory address locations at which a program executes.
Similar to ASLR, various fine-grained randomization techniques~\cite{MEMORYEXPLOIT,Hiser:2012:IWM:2310656.2310723,Wartell:2012:BSS:2382196.2382216} have been developed to defend against control-flow attacks. Apart from the difficulty of applying some of these techniques at the OS level~\cite{ASRMINIX}, none of these techniques alone can effectively stop JIT ROP
attacks~\cite{JITROP}.

\subsection{Position-Independent Code (PIC)}
\label{sec:pic}
For better ASLR support, user space code is typically compiled as
position-indepe\-ndent executables (PIE) and/or shared libraries.
Such code uses the ``RIP-relative'' addressing mode~\cite{AMD,INTEL}
in x86-64, where 32-bit offsets
are added to the instruction pointer, effectively allowing the program to
execute anywhere in the 64-bit virtual address space.\footnote{AMD
CPUs currently restrict their virtual addresses to 48 bits. Intel has recently extended virtual addresses to 57 bits with 5-level paging. Virtual addresses can be further extended to 64 bits in the future.}
Moreover, relative addresses, even if leaked, do not directly
reveal the absolute addresses needed for ROP attacks.
This widely advocated~\cite{LIBCRET} model
is steadily gaining popularity in Linux distributions~\cite{UBUNTUPIE} for user space protection.

The Linux kernel itself and its modules, however, do not employ the position-independent model.
While there exists a preliminary effort~\cite{LINUXPIE} to compile a kernel
image as a PIE, it falls short on addressing the same problem for kernel modules.
Since modules still use 32-bit KASLR,~\cite{LINUXPIE} currently extends the kernel's
KASLR range to 3GB only, which does not make a significant practical difference.
Moreover, most of the code nowadays is compiled outside of the kernel image,
e.g., Ubuntu 18.04's out-of-the-box kernel has over 5000 modules.

Our work extends the position-independent model for kernel modules.
Thus, it complements the existing PIE patch so that the
entire 64-bit range can be used for KASLR. Moreover, our design
enables the kernel and the modules to lie any distance apart from
each other, i.e., they do not necessarily need to be placed within
a $\pm 2$GB range of each other.

\subsection{Meltdown}

Meltdown (CVE-2017-5754) is a CPU attack aimed at exploiting the kernel portion of page tables. Fortunately, Linux's KPTI mitigation based on the KAISER~\cite{KAISER} page table isolation is fully transparent to the user. This mitigation does not impact any of our design choices.

\subsection{Spectre-V2}

Spectre-V2 (CVE-2017-5715) is an attack aimed at exploiting CPU
vulnerabilities based
on speculative execution and branch prediction~\cite{SPECTRE}. This attack affects
most existing CPUs, even outside of the x86-64 realm.
The system is vulnerable due to indirect
\textsc{call} (or \textsc{jmp}) instructions.

Linux uses a mitigation~\cite{RETPOLINE} which replaces indirect function
calls with direct calls to special \textit{retpoline thunks}, which, in turn,
use a workaround based on a \textsc{ret} instruction trampoline to prevent
speculative execution. The Linux kernel implements retpoline through special macros such as
\textsc{CALL\_NOSPEC} and \textsc{JMP\_NOSPEC} for assembly
code and relies on compiler support for C.

As x86-64 does not support 64-bit offsets for direct calls,
indirect calls must be used for 64-bit addresses.
It is crucial to reduce the number of indirect calls since
they use retpolines.

\subsection{GOT and PLT}

ELF shared (dynamic) libraries~\cite{X86ABI} provide a special
mechanism for handling external symbols. Since external symbols are unknown
at compile time, compilers rely on a \textit{global offset table}
(GOT) to retain this information. Instead of specifying addresses directly,
the compiled code needs to retrieve addresses from the corresponding GOT entries.

GOT is also important for other reasons. Dynamic libraries need to
be shared across multiple processes that can use different virtual
address ranges. Thus, PIC is typically preferred
(or, in the case of x86-64, required) for shared libraries to avoid the 
copying overhead. Since absolute references to code and data are
not identical in different processes, GOT encapsulates these addresses
so that only GOT needs to be updated when sharing the code with a different process.
Another reason to use GOT in x86-64 is for
storing complete 64-bit addresses as most instructions support only
32-bit displacements.

We found GOT to be particularly useful for continuous re\hyp{}randomization.
Although the kernel uses a single address space, and shared libraries are not directly useful in the kernel, we still want to efficiently support
multiple mappings to the same code due to the ongoing re\hyp{}randomization.
GOT allows to do so without
modifying the underlying code. Although not directly provisioned
by ELF shared libraries, we create multiple GOTs for different
purposes within the same module to facilitate continuous re\hyp{}randomization.

Dynamic libraries also use \textit{procedure linkage tables} (PLTs) to
transparently interpose on exported functions (e.g., a custom
\textit{malloc(2)} can interpose on \textit{libc}). PLT is also
used for lazy binding by dynamic linker trampolines.
Although PLT is typically useless for the kernel, we use it when the retpoline
mitigation is required
for better code efficiency (Section~\ref{impl:format}).

\subsection{Kernel Re-randomization Challenges}

Shuffler~\cite{SHUFFLER}, CodeArmor~\cite{CODEARMOR}, and TASR~\cite{TASR} previously explored re\hyp{}randomization for user space. Although Shuffler's, CodeArmor's, and TASR's goals are partially aligned with ours, the challenges are not identical. User-space
techniques do not handle low-level code such as system calls, interrupt
handlers, etc.
Kernel code is also often written to be agnostic to threads that use it:
function calls can go all the way from system calls initiated by
different user processes. Moreover, Shuffler benefits
from existent rich support for PIC
in user space. Early attempts~\cite{ASRMINIX} to solve a similar problem for more componentized OS designs such as MINIX~\cite{MINIX} pointed out many similar
challenges of implementing ASLR and re\hyp{}randomization
in OS environments.

We also take a different approach for performing re\hyp{}randomization.
Unlike Shuffler and CodeArmor, Adelie avoids \textit{binary-level} transformation. As code is available for
the kernel and most of its modules, it makes sense to
have a solution that requires a relatively small number of changes while
benefiting from \textit{source code} access. Shuffler also has
limitations such as requiring an executable and its associated libraries
to be within $\pm 2$GB from each other, effectively transforming
dynamically-linked applications into monolithic, statically-linked binaries. Similar limitations exist for CodeArmor, which does not benefit from PIC.
By design, CodeArmor chose to transform PIC to absolute-address code (i.e., subject to the 2GB limit unless executables are compiled with costly \textsc{mcmodel=large}~\cite{X86ABI}) to simplify its re-randomization process due to its extra layer of indirection.

\subsection{Driver VMs}
Xen \emph{driver domains}, unprivileged guest VMs that run device drivers,
are used in both desktop~\cite{QUBEOS} and enterprise~\cite{SECUREVIEW} setups.\footnote{We integrated Adelie into a similar enterprise system~\cite{SAVIOR} for better security.} Driver VMs isolate potentially vulnerable/malicious drivers
(or devices) from the privileged VM (Dom0). They also offload Dom0,
thereby increasing performance. Driver VMs have direct access to the underlying hardware~\cite{PCIPASSTHRU} and can be used for both networking and storage by using special \emph{paravirtualized} I/O drivers for communication.

Adelie is by no means limited to driver VMs, but they exemplify a
use case where driver re-randomization complements already
existent strong protections against vulnerable code.

\section{Design}

In this section, we discuss the challenges that pertain to
transforming modules to PIC as
well as module re\hyp{}randomization.

Our design is Linux-centric, but other OSs that rely on ELF modules (e.g., BSD) can implement similar mechanisms by making similar changes in their respective kernels. Non-ELF systems can adopt certain aspects of our approach (e.g., Windows also extensively uses the ``RIP-relative'' mode in x86-64) but may need some adaptations or substitutes for GOT and PLT.

\subsection{Threat Model}

Adelie aims to protect the Linux kernel from all known code reuse
attacks that are applicable to kernel space, i.e., both traditional
ROP and JIT ROP attacks.
In Section~\ref{sec:analysis}, we discuss why Adelie provides protection for the entire
kernel ecosystem even though we typically want to re-randomize vulnerable modules only,
regardless of whether ROP gadgets are available elsewhere (e.g., in the core kernel).
We only focus on x86-64 and assume that all existent protections
such as the NX bit are enabled by the kernel.
Although not required, users can leverage the stack protector mechanism used by
the kernel~\cite{STACKPROTECTOR} for even stronger security guarantees.

We make the following assumptions:
\begin{compactenum}[1.]
\item Attackers perform traditional control-flow attacks based on 
code reuse. NX already prevents code injection attacks.

\item Attackers primarily target kernel modules (drivers), which
are more likely to have vulnerabilities in the first place.

\item We do not specifically aim to protect from  non control-flow
attacks~\cite{DATAFLOW}. However, since we re-randomize the static
data layout, such attacks become more challenging.

\item Privilege escalation by data modification (e.g., \textit{struct cred}) requires some vulnerability. Short of trivial kernel bugs, an attacker is likely to use vulnerable modules (drivers). The attacker attempts code reuse attacks by constructing a ROP chain which modifies data (e.g., by leveraging \textit{ioctl(2)} from user space).

\item Hijacking of kernel page tables is infeasible since
the core kernel itself is unlikely to be vulnerable.

\end{compactenum}

\subsection{Goals for Continuous ASLR}

We define the following main goals for Adelie:

\begin{description}[style=unboxed,leftmargin=0cm]

\item[Generality]
We aim to transform \emph{all} modules to the 64-bit KASLR model.
For continuous KASLR and stack re\hyp{}randomization, we aim to
make the process as automatic as possible.

\item[Performance]
For re\hyp{}randomization, the number of
absolute addresses must be minimized. Since address layout
changes frequently, we aim to avoid the cost of copying code and data.

\item[Entropy]
Every module should be any distance apart from other modules and
the kernel. Thus, leakage in one kernel module or the kernel does not
automatically reveal the larger picture of the address space layout.

\item[Security]
We aim to protect against code reuse attacks by minimizing the time
duration during which module addresses remain valid. We also encrypt return addresses with a key that is continuously re-randomized.

\end{description}

\architecture

\subsection{Extending KASLR}

Our first step is to extend Linux's KASLR support. Since position-independent code can reside anywhere, PIC extends KASLR to 64 bits while avoiding costly (absolute-address) models such as \textsc{mcmodel=large}~\cite{X86ABI}. By fully converting all modules to the PIC model, we can also attain performance benefits outlined in our goals by eliminating code modification during re-randomization.

As discussed in Section~\ref{sec:pic}, there exists preliminary
support for compiling the Linux kernel as a position-independent executable (PIE). 
It can be considered analogous to running position-independent executables in user space. 

PIE, for the most part, only changes the absolute address mode to the ``RIP-relative''
mode. All global variables are still assumed to be within
$\pm 2$GB reach.\footnote{PIE optimizes external symbols because they can be allocated
within $\pm 2$GB from the executable image even if imported from shared libraries.}
Thus, a kernel PIE avoids the global offset table (GOT).
The procedure linkage table (PLT) is not used by kernel PIEs since the kernel does not directly call any outside function (as opposed to user space PIEs which call functions from shared libraries).

Kernel modules, however, cannot use PIE because they can be placed any
distance apart from each other and the kernel. Instead, Adelie uses a more
general PIC model from shared libraries with GOT and PLT support.
However, as discussed further in Section~\ref{impl:format},
we do not convert modules to shared libraries but use position-independent
relocatable objects.

\subsection{Continuous Module Re-randomization}

\label{sec:design}

The most vulnerable kernel code
(i.e., device drivers, certain kernel libraries, etc.) can be compiled as
modules and subsequently re\hyp{}randomized.
The core kernel and non-vulnerable modules do not have to be re-randomized
because stack address encryption, as described below, defends
against \textit{any} ROP gadgets. Gadgets
are also likely to come from code which is more predictable to an attacker, such as device drivers, for which we use continuous ASLR.
We now describe how we achieve the goal of efficient
run-time re\hyp{}randomization of kernel modules.

\subsubsection*{Zero-copying Mechanism}
One critical aspect of our work is that, unlike Shuffler~\cite{SHUFFLER}, we
completely avoid copying code and static data while re-randomizing addresses. Similarly to CodeArmor~\cite{CODEARMOR}, we remap existing pages to new addresses but favor a more fine-grained memory reclamation technique to quiescent state-based reclamation (QSBR)~\cite{epoch2}.
In Figure~\ref{fig:architecture}, we demonstrate the high-level principle.
Initially, at instant 1, both the kernel and the modules are in some randomly chosen
address space, any distance apart from each other. This is achieved using
our extended 64-bit KASLR. Periodically, module locations
are re-randomized by a special \textit{randomizer} kernel thread. This thread
creates new mappings, as shown at instant 2. Finally, when old regions
are no longer used, we unmap them. In this process, no copying is actually
made; we simply create new page table entries that point to the same
physical memory locations.

\subsubsection*{Module Organization}
Re-randomizable modules consist of two logical
parts: the movable (most of the code and data) and immovable, which
mostly implements
glue code for the kernel. While for convenience the immovable part is placed
in the module, it can be viewed more as an integral part of the kernel.
In Figure~\ref{fig:design}, we show a typical
module layout. Because the immovable part can be any distance away
from the movable part, we maintain two different sets of GOTs.
Each set of GOTs stays within $\pm 2$GB of the corresponding
logical part. Each part can reference global (kernel or fixed) symbols
or module-local symbols. Only module-local symbols need to be updated
when modules move.
For this reason, each set contains two GOTs for fixed and local
addresses, respectively.
We only update (i.e., reallocate) local GOTs when moving modules. Moreover, we minimize the 
number of entries in local GOTs (see Section~\ref{impl:format}).
In the figure, local GOT entries always point to the movable part. Fixed
GOT entries are either referring to the kernel or to the immovable part of the
module.

\subsubsection*{Controlling Address Space Lifetime}
It is crucial to unmap pages from previously used virtual addresses in a timely manner, so that ROP gadget addresses quickly become obsolete and useless for
an attacker. However, kernel code is quite complex in the way
functions from modules are called. A long chain of functions
can be called, eventually leading to some user
space thread making a system call. It is infeasible
to use existing techniques of stack unwinding (e.g., as in Shuffler~\cite{SHUFFLER})
while pausing the execution of the entire kernel.

Instead, we opted to use delayed unmapping. We let pending calls finish
using previously obtained virtual addresses.
Any call that follows re\hyp{}randomization must use new virtual
addresses. As soon as the last pending call completes, the previous
address range is immediately unmapped. Since almost all
kernel space calls should be relatively quick (or, otherwise it 
would indicate bugs in the code), unmapping is not
delayed substantially enough for an attacker to benefit from it.

\wrapper

The challenge is to track pending calls with little
overhead and in a scalable manner.
We use the Hyaline reclamation scheme~\cite{Hyaline,hyalineFULL}, which is similar to epoch-based reclamation (EBR)~\cite{epoch1,epoch2}. Hyaline's performance is very similar to that of EBR, but Hyaline enables much easier integration into the Linux kernel since it is context-agnostic and does not make any assumptions about how threads are managed. Both Hyaline and EBR largely solve the same problem of efficient, optimistic memory access to blocks that are concurrently being deallocated by other threads. We enclose operations that
access potentially disappearing memory blocks with \verb|mr_start| and \verb|mr_finish|.
These operations postpone memory reclamation until after all pending calls (i.e., those that
called \verb|mr_start|) execute \verb|mr_finish|.
Memory blocks are not deallocated directly, instead they are first \textit{retired} with
a special \verb|mr_retire| operation. The deallocation takes place only after pending calls complete.

As in CodeArmor, unmapping can be delayed, but that does not appear to be a practical issue for modules we tested. Linux also has built-in mechanisms to detect calls that block for too long. Blocking is more likely to manifest when using softirqs/workqueues. However, softirqs/workqueues do not require \verb|mr_finish| to wait until the request is completed, and the re-randomization routine will only need to modify the function handler address. Only inside the actual handler (when scheduled), do we need to call \verb|mr_start|/\verb|mr_finish| again. Blocking, in general, can be more challenging but typically can be handled easily, i.e., by inserting \verb|mr_start|/\verb|mr_finish| and controlling the state around these calls manually.

\subsubsection*{Function Wrapping}
Externally accessible functions from re-random\-izable modules are specially
wrapped for two reasons: (1) a module moves in the
address space while the kernel retains absolute addresses to the module code;
by placing a wrapper function into the immovable part of the module,
the kernel can reference that wrapper function instead; since we create the
local GOT in the immovable part as well, there is an easy way to update
references to the original function which is placed into the
movable part.
(2) lifetime control: almost all
function calls initiated from the outside must be protected by
the special \verb|mr_start| and \verb|mr_finish| calls from
the memory reclamation algorithm.

In Figure~\ref{fig:wrapper}, we show how an externally accessible
function is transformed by renaming it and placing a wrapper with
the original name into the immovable part. The kernel
references the wrapper rather than the original function.
A special compiler plugin (Section~\ref{sec:implementation})
automatically
wraps functions.

\subsubsection*{Stacks}
\label{ref:stackreassign}
We continuously re-randomize stacks, so that
an attacker cannot hijack the control-flow pointers placed on
the stack. Changing stacks is not a trivial task as the kernel
maintains multiple stacks for user- and kernel space.
Each thread needs its own stack, and stack re-randomization must
be fast.

We maintain a per-CPU lock-free (LIFO) list of stacks. We substitute stacks at the beginning of the function wrapper by dequeuing the head of the per-CPU list. New stacks are allocated on demand as needed.
In Figure~\ref{fig:stack}, we show \verb|get_new_stack|
and \verb|return_old_stack| code used in wrapper functions.
Old \verb|%rsp| is saved into \verb|%rbp|, and \verb|%rsp| is replaced with the newly dequeued stack. When exiting, the stack is restored and returned
to the list.
Since each CPU has its own list, the contention is low and can only occur due to the re-randomizer thread deleting old stacks.
The re-randomizer thread generates new LIFO lists for each CPU.
Old list heads are atomically replaced with new heads. The old lists
are garbage collected and freed (when it is safe).

Return addresses can potentially be hijacked by
jumping to non-rerandomizable code. Since they are
not strictly controlled as other non-rerandomizable pointers (Section~\ref{sec:analysis}), it is crucial to encrypt them.
For that reason, the \textit{prologue} and \textit{epilogue} of each function in
re-randomizable modules are identically modified to encrypt and decrypt
(XOR with a key) the return address as shown in Figure~\ref{fig:stack}.
Static functions occasionally use custom calling conventions, and we
cannot use \verb|%r11| as a scratch register.
Since frame pointers are also set in the prologue and epilogue, we
recycle \verb|%rbp|; we also optimize out \verb|push| or \verb|pop|, respectively.
In case of \verb|%r11|, we clear the register to avoid
accidental key leakage.
The encryption key is randomly generated and
stored in the local GOT, which gets reallocated during
re-randomization. The key changes every re-randomization.
Even if an attacker manages somehow to read
the encryption key from the GOT, it becomes obsolete every time the module
is re-randomized. Moreover, the absolute address of the local GOT also
changes due to re-randomization.

Since up to six arguments are passed through registers in x86-64
(we have not discovered any function to wrap with $> 6$ arguments),
we simply replace stack pointers for a duration of the call. If any arguments have pointers to stack data, they will simply reference the original stack. However, all functions that are executed
from the module will use the new stack.

\section{Implementation}
\label{sec:implementation}

The overall implementation effort is reasonable.
The changes for PIC modules ($\approx$ 727 LoC) assume that the Linux kernel
is already patched by the existent PIE patch~\cite{LINUXPIE}, which is related
to our change but only extends KASLR for the kernel itself.
With our change, \emph{all} ($\approx$ 5000) modules use 64-bit
(basic) KASLR. Our system is stable and self-hostable on
different machines (with Ubuntu 18.04's default configuration).
We confirmed that our system can still work in specialized scenarios,
e.g., when running Linux as Dom0 in Xen.

With respect to PIC adaptation, obstacles were primarily in assembly files, non-standard calling conventions, C macros with inline assembly, hypervisor interactions, and special calls (e.g., exception handlers).

Re-randomization of modules is implemented using a common part for all modules ($\approx$ 2815 LoC).
Using our GCC plugin ($\approx$ 1400 LoC), we automatically modified and tested
network (E1000E, E1000, ENA),
storage (NVMe), USB 3.0 (xHCI), and file system (FUSE, ext4) drivers.
Although it is infeasible to rigorously test every single device
driver, our approach is more or less general; we also confirmed that
the plugin (at least) successfully compiles \emph{all} kernel modules.

\begin{figure}[ht!]
\includegraphics[width=\columnwidth]{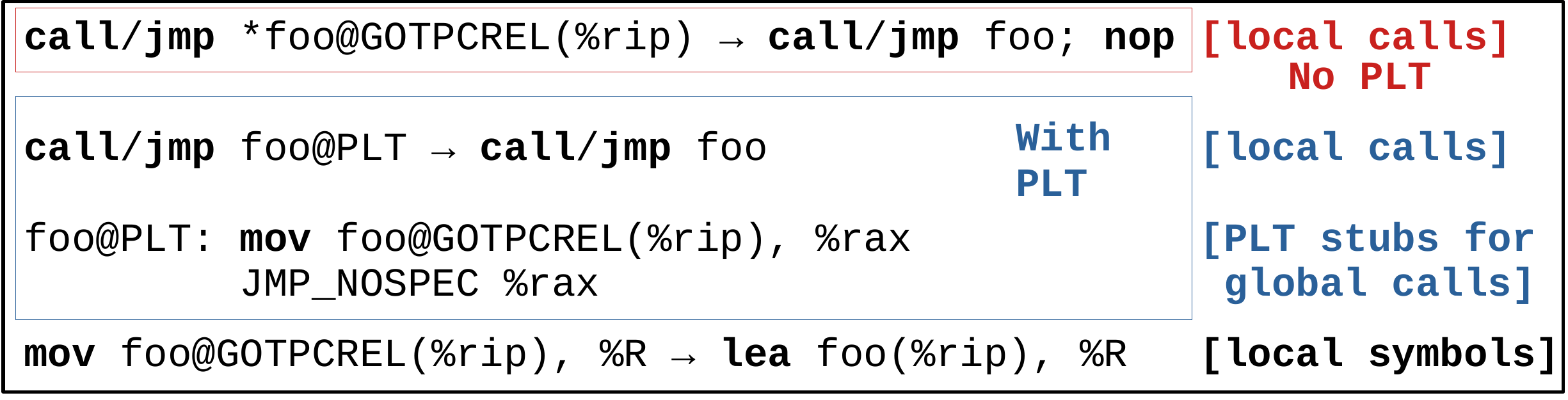}
\caption{Summary of run-time patching.}
\label{fig:patching}
\end{figure}

Our GCC plugin is \emph{only} needed for continuous ASLR. It detects and wraps all functions and variables that are
exposed to the kernel by using predefined macros and modifying the compiler output accordingly. Similarly, for all functions inside modules, the plugin automatically applies \textit{prologue} and \textit{epilogue} changes as discussed above.

Our plugin makes its best efforts with respect to the C-language semantics. Although exceptions from typical rules are certainly possible, they should be rare in practice since programmers generally follow good-style Linux guidelines in the code. Also, if problems with symbols arise, it should be very easy to detect external (non-rerandomizable, kernel) addresses since they are marked as U (undefined) in the corresponding module symbol.

\subsection{Executable Format}
\label{impl:format}

We kept the existing relocatable format and adapted it for PIC.
This enables larger flexibility
when handling GOT and PLT, as relocations are only finalized at run-time.
Whereas shared libraries have just one GOT, we allocate four GOTs: two tables for the movable and two tables for the immovable module parts. 
One table from each pair is used for the imported kernel (or the immovable part) addresses, and the other one is used for module-local addresses which are periodically changed.

Since the location of symbols is often unknown, the compiler generates
code which can be suboptimal for local symbols and calls.
To optimize code, we patch it at run-time (see Figure~\ref{fig:patching}).

If the retpoline mitigation~\cite{RETPOLINE}
is disabled (for newer CPUs), we avoid PLT stubs by inlining them when
compiling modules.
When loading modules, we patch relocations to optimize instructions that
use module-local symbols because they are known to be within the $\pm 2$GB range
from the instruction pointer.
Since a direct \textsc{call}/\textsc{jmp} is one byte shorter
than its indirect counterpart, we pad it with \textsc{nop}.

\begin{verbbox}[\footnotesize]
\end{verbbox}

When retpoline is enabled, the above-described approach creates
obstacles for local call optimizations as corresponding retpoline-based calls
use longer sequences of instructions, clobber registers, etc.
To support this case efficiently, we create PLT stubs using
Linux's JMP\_NOSPEC.\footnote{Linux
occasionally uses non-standard calling conventions. \theverbbox\ seems
to be the only safe volatile register (also used for return values) across
almost all conventions. We handle a few exceptions separately.}
Local calls are optimized by just eliminating respective PLT stubs.

Linux keeps 32-bit relative
addresses for handler functions
when generating exception tables. Since the number of
global handlers is small, we avoided intrusive changes
by using PLT stubs for exception handlers even in the
non-retpoline case. (While implementing this change, we discovered a
subtle bug in relocation handling by the GNU assembler, which was subsequently resolved.)

Variables are also optimized.
By default, the compiler generates code that uses GOT
to retrieve variable addresses; we patch relocations to use the \textsc{lea} instruction for local symbols.

Aside from direct performance benefits, the aforementioned optimizations
substantially reduce the total number of GOT and PLT entries, thereby
reducing the risk of leaking absolute addresses to an attacker.
Moreover, when performing continuous re\hyp{}randomization, we have
to change addresses in GOT entries for local symbols. Thus, we only
need to change a few, thereby reducing the added cost of re\hyp{}randomization.

We write-protect pages with GOT/PLT entries after initialization so that they
cannot be overwritten by an attacker.

\subsection{Module Re-randomization}

A special process takes place when loading re-random\-izable modules. First,
we identify sections that belong to movable and immovable parts and allocate
them separately. We only place ``.fixed.text'' (the section that contains function wrappers)
and ``.rodata'' in the immovable part. The reason why we currently place 
read-only module data in the immovable part is mostly to avoid excessive
changes in the module because certain constants and string literals
can be directly passed to the kernel, and we do not want to re-randomize their
location.

For re-randomizable modules, we use four different GOTs. Based on whether the code is an immovable or movable part and symbol locality, we choose one of
these GOTs when generating a GOT entry. Occasionally, two tables
will contain duplicates of the same symbol (e.g., a local GOT entry
from the immovable part points to the same address as a GOT entry
from the movable part). We use separate GOTs for movable and
immovable parts because these parts can be any distance 
away from each other, 
while GOTs must always be placed within $\pm 2$GB reach from \%rip due to the
``RIP-relative'' address mode.

During module re\hyp{}randomization, a special kernel thread (``\emph{re-randomizer}'') periodically performs the following steps.
First, a new virtual address space map is created; it maps to the same physical addresses
as an old map. New local GOTs are allocated for both movable and immovable
parts of the module. All entries from the previous GOTs are adjusted to point to the 
new memory address space when creating the new GOTs.
Corresponding GOT pages in the new address space are remapped to point to the new GOTs.
Subsequently, the re-randomizer thread calls a special function from the module
to update its run-time function pointers (needed for some modules). Finally, the re-randomizer
thread calls \verb|mr_retire| from the memory reclamation algorithm to request unmapping.
After that, the re-randomizer thread sleeps for the specified re\hyp{}randomization period, and then
repeats the entire process.
The memory reclamation algorithm unmaps the previous address space when all pending
calls complete.

\subsection{Limitations}
Our re-randomization approach is relatively coarse-grained. Although this may certainly have drawbacks, Adelie's current implementation makes the very first step towards continuous module re-randomization. If more fine-grained re-randomization is desired in the future, it can be attained at least at a function-granularity: we would need to create separate GOT tables per each function or a group of functions. This approach is especially useful for more fine-grained re-randomization of larger modules. Although each module can still be self-sufficient in terms of ROP gadgets, frequent address changes prevent the attacker from performing a successful attack.

Frequent address space remapping may contribute to re-random\-ization costs. More specifically, TLB needs to be flushed more frequently after page table updates. We did not find that to be a significant problem in Section~\ref{sec:eval} for re-randomization periods that we used. The TLB cost is also somewhat unavoidable for continuous re-randomization even for other alternatives (e.g., simple copying), where the page table would still have to be updated frequently due to the 64-bit address space being very sparse.

\begin{table}
\caption{Server and client systems.}
\label{tbl:system}
\begin{center}
\setlength\tabcolsep{2.1pt}
\small
\begin{tabular}{ l l l }
\toprule
 & \textbf{Server (for Evaluation)} & \textbf{Load Generator} \\
\midrule
CPUs & Xeon Silver 4114 2.20GHz & Core i7 4770 3.40GHz \\
Cores & 2x10, no HyperThreading  & 1x4, no HyperThreading \\
L1/L2 cache & 64 / 1024 KB per core & 64 / 256 KB per core \\
L3 cache & 14080 KB & 8192 KB \\
Memory & 96 GB & 16 GB \\
Network & Intel E1000E 1GbE & Intel E1000E 1GbE \\
Storage & Samsung 970 EVO NVMe & Samsung 860 EVO SSD\\
USB 3.0 & Intel C620 xHCI & N/A \\
\bottomrule
\end{tabular}
\end{center}
\end{table}

\section{Experimental Evaluation}

\label{sec:eval}

Our primary goal is to determine Adelie's performance overheads (if any). We use various microbenchmarks
to evaluate overall system
performance when running position-independent modules.
We also use macrobenchmarks including ApacheBench~\cite{APACHEBENCH} and SysBench/OLTP (mySQL)~\cite{web:sysbench} to evaluate the performance of real-life server applications, specifically when using continuous re\hyp{}randomization.
Finally, we considered the SPEC CPU benchmarks, but since they are CPU intensive, we did not 
observe significant differences, and consequently we do not present SPEC CPU results.

\begin{figure}[ht!]
\centering
\subfloat[Module Size]{
\includegraphics[width=.47\columnwidth]{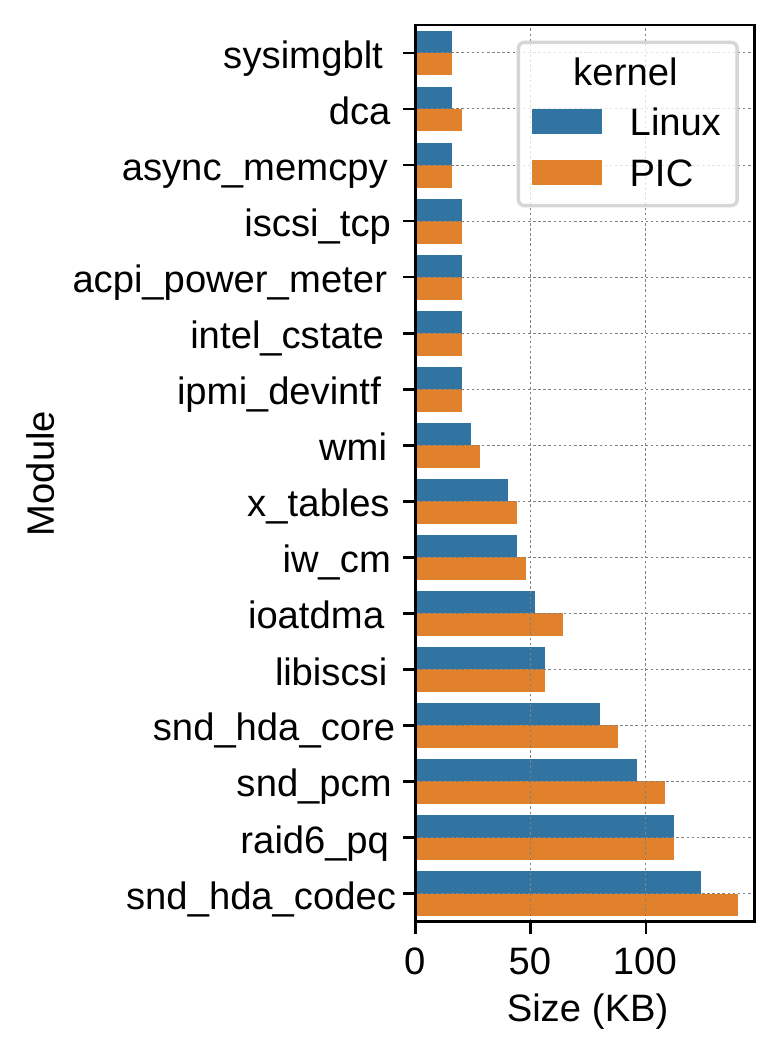}
\label{fig:modulesiszes}
}
\subfloat[Microbenchmark]{
\includegraphics[width=.47\columnwidth]{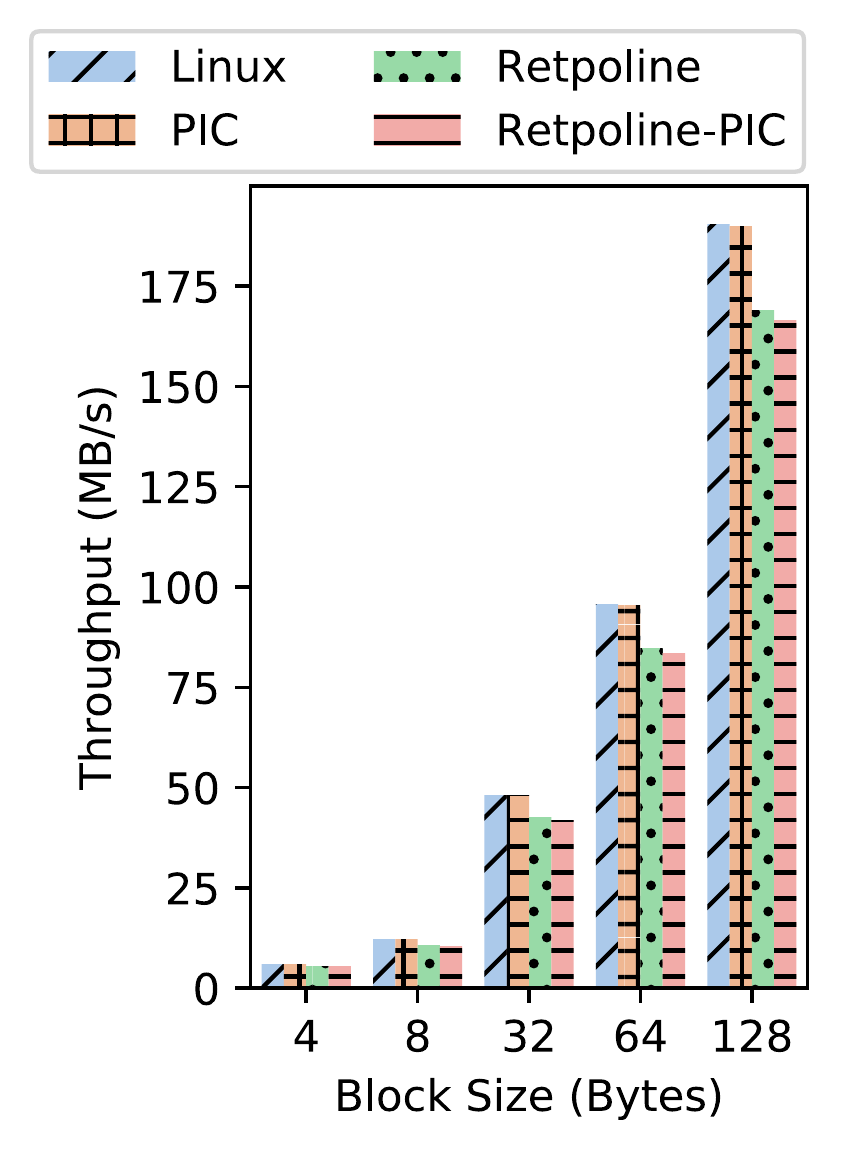}
\label{fig:file_micro}
}
\\
\subfloat[Sysbench (Cached)]{
\includegraphics[width=.47\columnwidth]{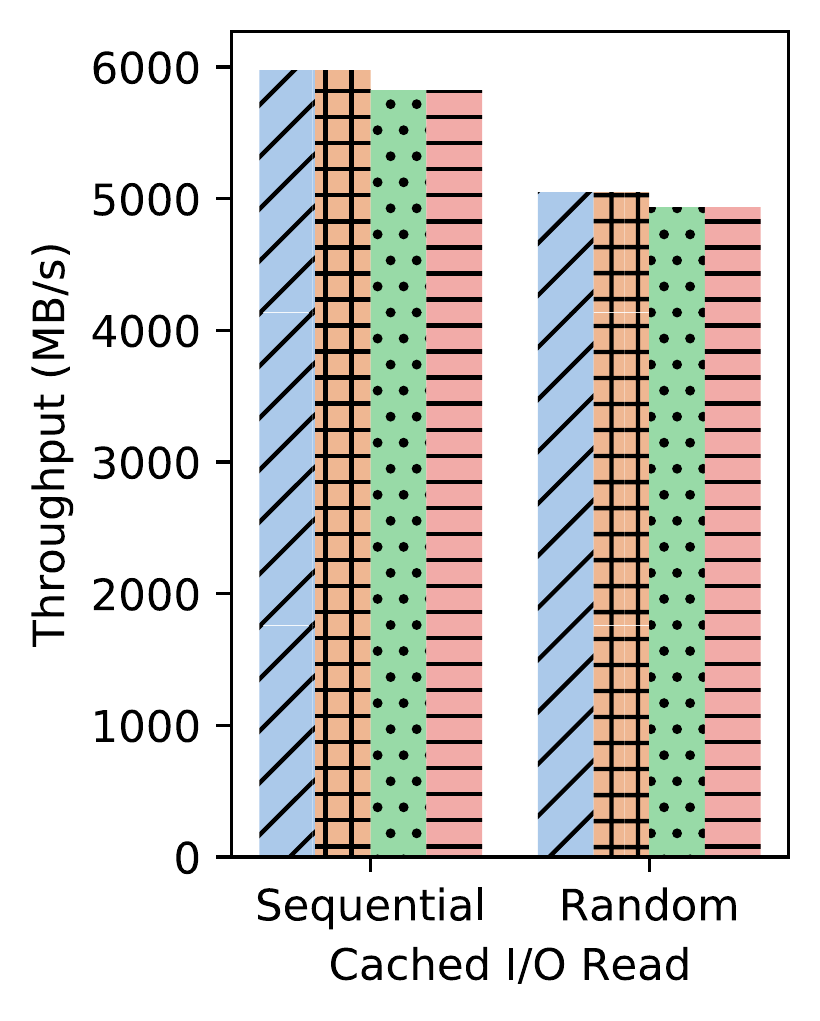}
\label{fig:sysbench_file}
}
\subfloat[Kernbench]{
\includegraphics[width=.47\columnwidth]{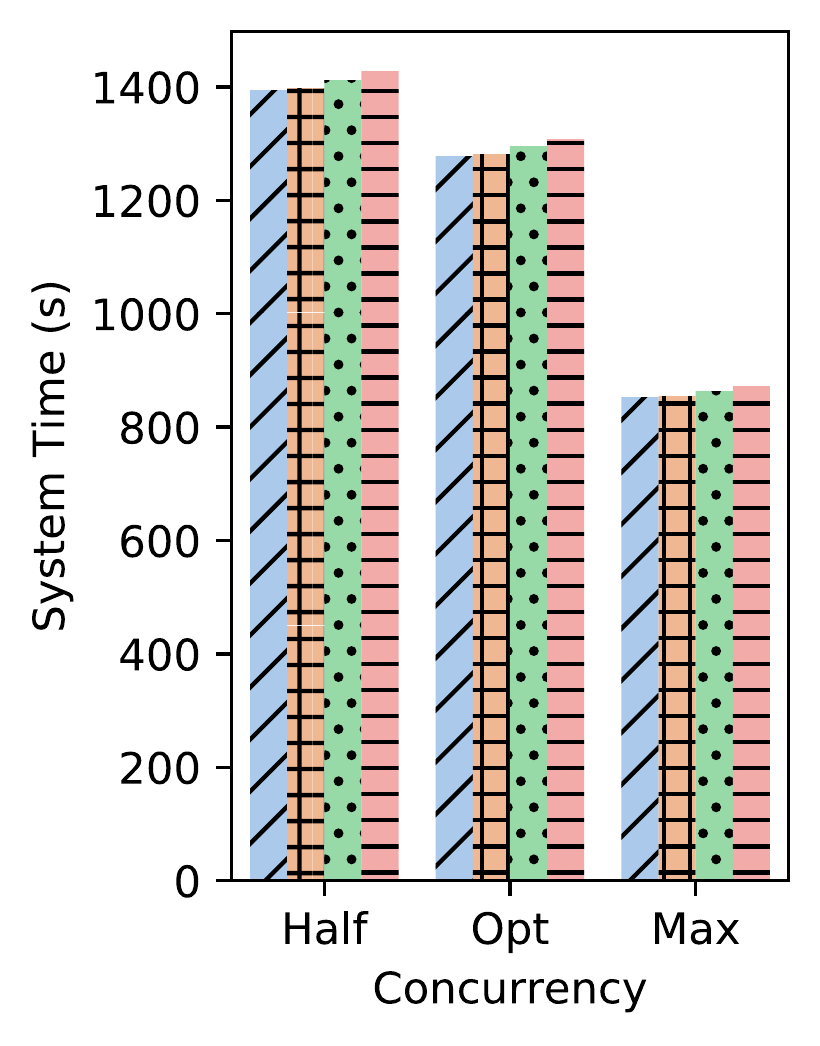}
\label{fig:kernbench}
}
\caption{PIC vs. non-PIC modules.}
\end{figure}

We evaluate each data point three times and present the average. Unless specified
otherwise, the standard deviation is $<0.5$\%.
As there is no other kernel re\hyp{}randomization system for Linux, we compare directly against original Linux.

In all of our tests, we use Ubuntu 18.04 with standard packages but a different Linux
kernel version (v5.0.4). In all test combinations, we use the
kernel with the default Ubuntu configuration, which compiles over 5000 various kernel modules. \emph{All} modules use the PIC model, presented in the paper.
We tested a fair amount of modules on various hardware to confirm that
PIC is bug-free across different modules.

Table~\ref{tbl:system} shows our experimental setup. For continuous re\hyp{}randomization, we evaluate widely used device drivers: Intel E1000E (network) and NVMe (storage). Additionally, we tested re-randomizable xHCI, FUSE (file systems in user space), and ext4 modules as an extra load. Finally, we successfully ran other network drivers including E1000 (used in VirtualBox) and ENA (used in Amazon AWS clouds).
This choice of re-randomizable modules is based on a reasonable assumption that device drivers are the most vulnerable components; we choose drivers from the most critical classes of drivers: network, storage, USB, and file systems.

\subsection{Position-Independent Modules}

We first measured PIC modules (our first contribution), their overall impact on system performance, and memory footprint.

In Figure~\ref{fig:modulesiszes}, we randomly selected modules of different sizes to demonstrate the memory footprint
related to the conversion to the PIC model.
The difference in memory footprint of the retpoline and non-retpoline modules is insignificant here; thus, we only present retpoline-enabled (PIC) and vanilla Linux modules. Overall, the overhead is negligible for all modules.

Next, we ran micro- and macro- benchmarks to evaluate the performance impact of PIC modules.
We use four setups: vanilla Linux without retpoline, vanilla Linux with retpoline, PIC modules without retpoline, and PIC modules with retpoline.
We used the default Ubuntu configuration in all tests.

\begin{figure}[ht!]
\centering
\includegraphics[width=\columnwidth]{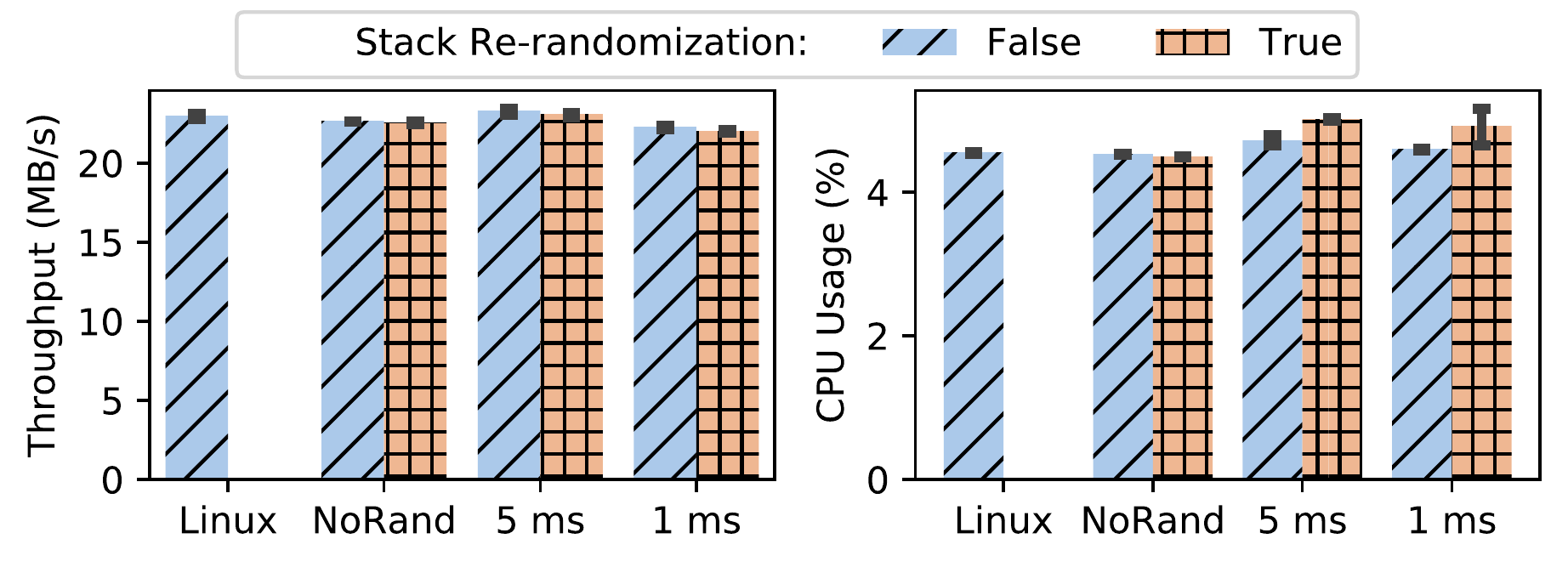}
\caption{NVMe read throughput.}
\label{fig:nvme}
\end{figure}

In Figure~\ref{fig:file_micro}, we ran our own microbenchmark which uses \verb|dd| to read files with varying block sizes. This test is CPU bound due to the use of the buffer cache. This experiment revealed the real impact of retpoline (a Spectre-V2 mitigation). Figure \ref{fig:file_micro} shows that without retpoline the performance of PIC and non-PIC is nearly identical. A slight performance hit of the PIC code (with enabled retpoline) is due to retpoline-safe indirect jumps to external functions in PLT stubs.

We used  the \verb|sysbench file_io| benchmark to measure the throughput on random and sequential reads. For this experiment, the files were cached in RAM to keep the results I/O invariant. The results in Figure~\ref{fig:sysbench_file} show that the performance of PIC-enabled and non-PIC systems is nearly identical.

\verb|Kernbench| is a CPU throughput benchmark that is often used to compare kernels. We recorded the time spent in kernel space at three levels of concurrency. The results in Figure~\ref{fig:kernbench} show no substantial difference
across different configurations.

Overall, PIC's cost is negligible even for enabled retpoline.

\subsection{Evaluation of Re-randomization}

To evaluate module re\hyp{}randomization, we use multiple
benchmarks. We first run experiments using typical I/O loads with ordinary device drivers. To estimate worst-case overheads, we also run a separate CPU bound test in Section~\ref{sec:cpubound} by designing a special driver which handles dummy IOCTL requests in a loop.

We use the retpoline-enabled kernel, as we previously already identified the cost of the retpoline mitigation.
For all tests, we present CPU usage across all 20 cores.
We evaluate up to \textit{five} different device drivers. We focus on the most critical classes of drivers, which are likely to be very attractive to an attacker.

To reliably evaluate the NVMe driver under re-randomization, we designed an experiment that minimizes the effects of I/O in the underlying hardware. We created our own benchmark that measures read throughput of a file stored on the NVMe storage. The file is opened with \verb|O_DIRECT| and \verb|O_SYNC| flags using the \verb|open| syscall, and a block size of 512 bytes is repeatedly read from the beginning of the file in a tight loop. The above-mentioned flags guarantee synchronous data transfer through the NVMe driver and prevent buffer caching
in the kernel. We read the same block over and over again to leverage NVMe's internal DRAM cache in an effort to minimize I/O wait time.
We measured the read throughput of the NVMe driver and recorded the CPU usage for the duration of the experiment.
We set different re\hyp{}randomization intervals (1 and 5 ms) and compare against vanilla Linux. Past works, e.g., Shuffler~\cite{SHUFFLER}, argued that even 50 ms is more than sufficient to prevent typical attacks, but we use much shorter intervals to reflect Adelie's performance in the event that intervals must be shortened in the future as the sophistication of security attacks grows. We also measure results when performing no re-randomization.
Apart from a slight increase in CPU usage (Figure~\ref{fig:nvme}), which is within the margin of error, the performance of NVMe storage remains largely unaffected by re-randomization. Enabling or disabling
stack re-randomization does not create much impact.

\begin{figure}[ht!]
\centering
\includegraphics[width=\columnwidth]{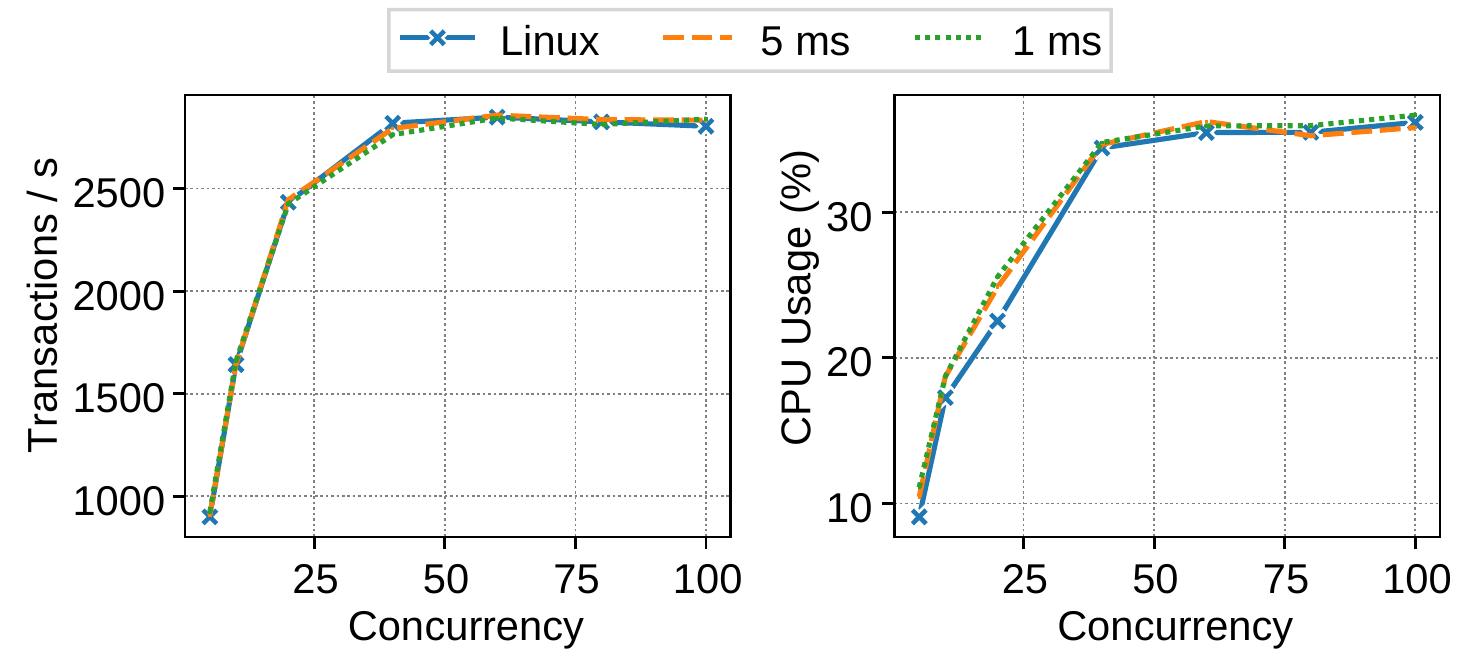}
\caption{mySQL with re-randomizable modules.}
\label{fig:sysbench_sql}
\end{figure}

\begin{figure}[ht!]
\centering
\includegraphics[width=\columnwidth]{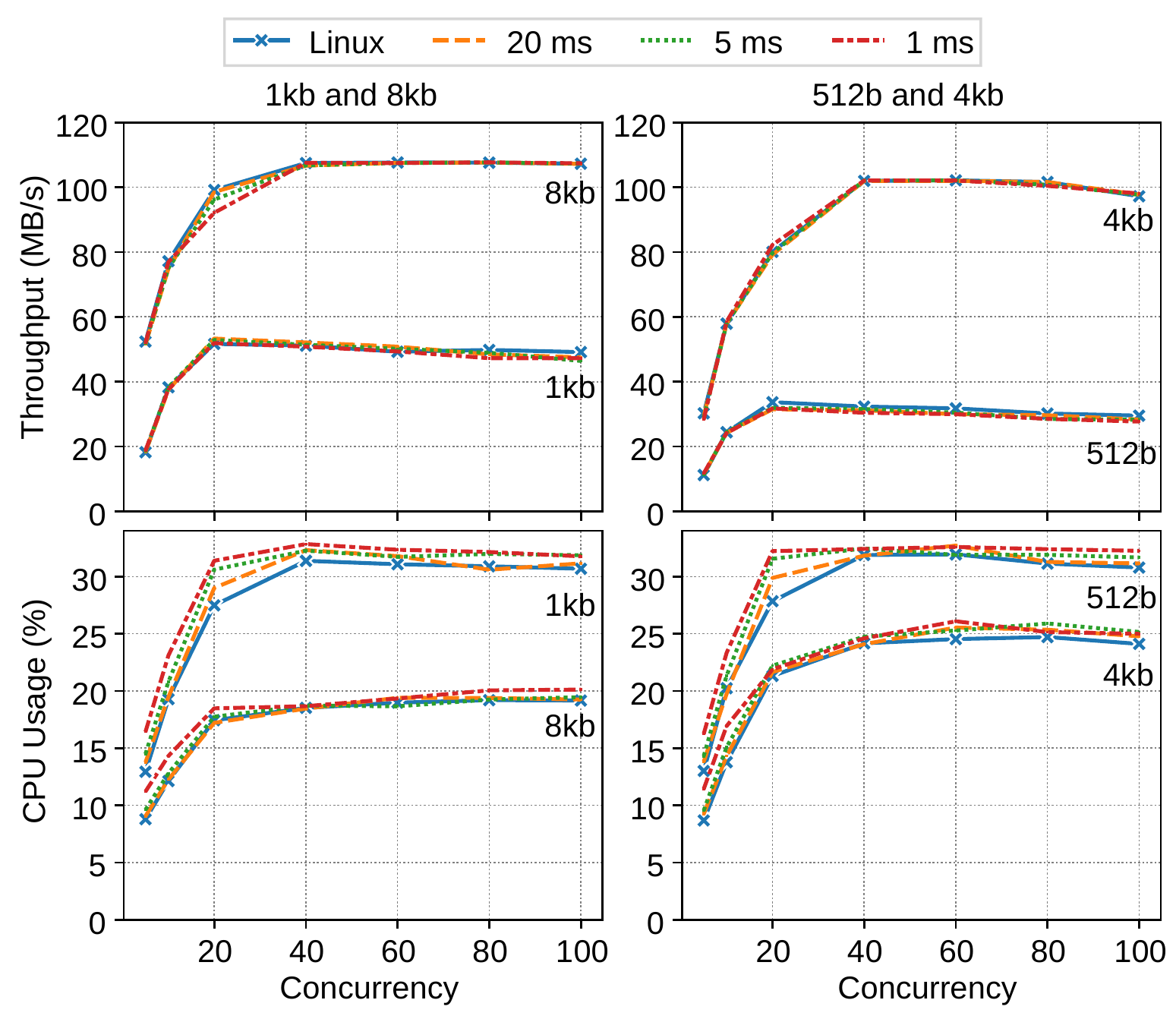}
\caption{ApacheBench with re-randomizable modules.}
\label{fig:network}
\end{figure}

We then measure network and storage intensive applications while
re-randomizing corresponding drivers. We use mySQL and Apache with their default configurations from Ubuntu. We run the corresponding macrobenchmarks from a client machine which is directly connected to the network adapter of our testbed (server).
We only present results with enabled stack re-randomization
as no substantial difference is observed otherwise.
We omit Linux with no re-randomization since its results
are almost identical to that of vanilla Linux.

We measure mySQL performance using \verb|sysbench| \verb|oltp| on a database comprising 10 tables with 1,000,000 rows of data each. The database is partially cached in memory and the experiment is conducted with varying levels of concurrency and different re-randomization periods. We re-randomize both
the E1000E and NVMe drivers. (Re-randomizing any of them alone yields
very similar results.) Figure~\ref{fig:sysbench_sql} shows that mySQL's rate
of transactions is identical for vanilla Linux, 1 ms, and 5 ms. The CPU usage
increases slightly ($<2$\%) prior to the concurrency saturation point.
The network throughput (going up to 110MB/s) is identical
across all configurations.

\begin{figure}[ht!]
\centering
\includegraphics[width=\columnwidth]{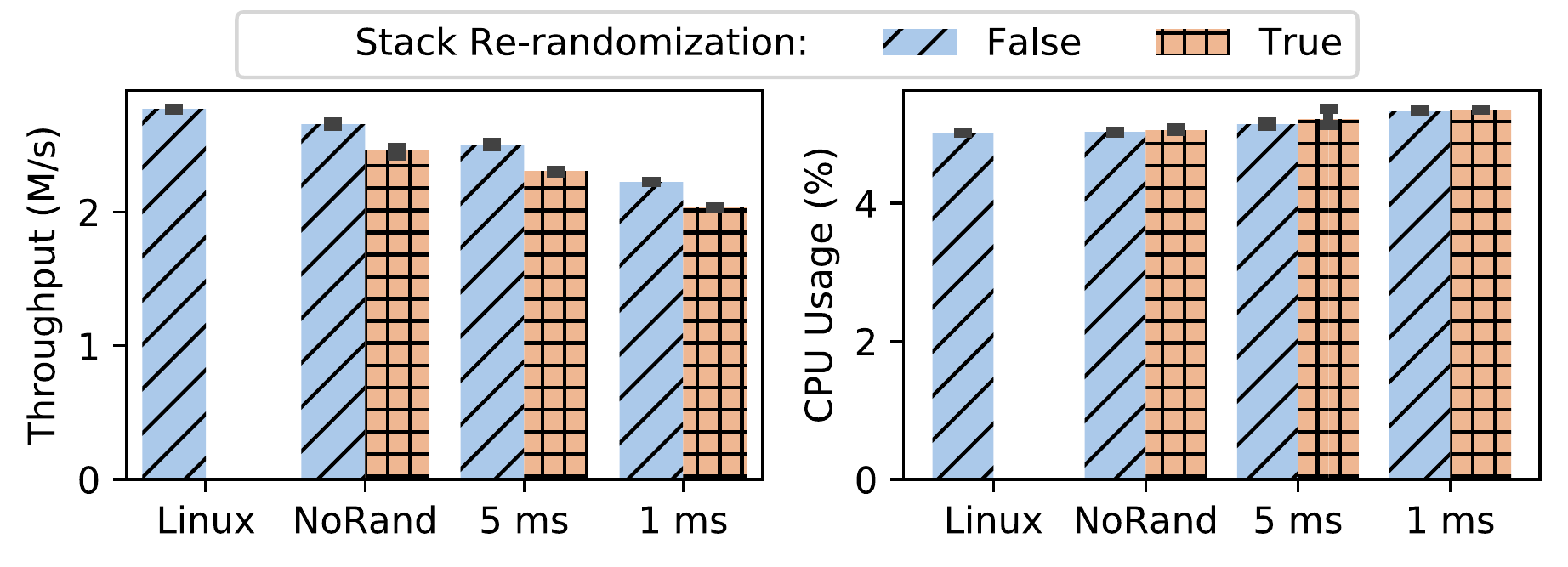}
\caption{IOCTL throughput.}
\label{fig:ioctl}
\end{figure}

In Figure~\ref{fig:network}, we present results for Apache using different block sizes (512B-8KB). We re-randomize several modules: E1000E, NVMe, FUSE, ext4, and xHCI drivers. In this test, the pressure is applied mostly to E1000E with occasional NVMe accesses. Other drivers are not on the critical path; they just provide an extra re-randomization load. Re-randomizing any driver alone yields very similar results. Smaller blocks put more stress on the system as the total number of system calls increases.
As Figure~\ref{fig:network} shows, re-randomization does not impact the overall system throughput. Re-randomization increases CPU usage
for smaller blocks ($\approx 2 \%$, except smaller concurrency).
We found that the CPU usage can be reduced by using 20 ms periods, which still provide sufficient security guarantees.

\subsection{CPU-bound test}
\label{sec:cpubound}
Most of our experiments on device drivers showed that re-random\-ization does not impact the device performance. This can be attributed to the fact that device drivers are I/O bound. The I/O wait time outweighs the CPU time by a large margin.

To test the extreme case of a CPU bounded driver, we designed a special microbenchmark.
We created a dummy device driver that implements a null \verb|ioctl| operation.
We repeatedly make the \verb|ioctl| syscall on the driver in a tight loop and measure the number of ioctl operations performed per second. Figure~\ref{fig:ioctl} shows the  throughput (in million operations per second) along with the corresponding CPU usage. This benchmark captures the impact of function wrappers and stack randomization. We found that function wrappers cause a performance drop of $\approx$~4\% and stack randomization causes an additional drop of $\approx$~6\% when compared to the original Linux.

\subsection{Scalability}
From the macrobenchmarks, we found that driver re-random\-ization does not
increase CPU usage significantly, which indicates great scalability.
Furthermore, adding an extra driver has very little performance impact, as we observed in OLTP and ApacheBench benchmarks, which simultaneously re-randomize several potentially vulnerable modules in the presented results.
Nonetheless, we wanted to get worst-case scenario estimates.
The CPU usage of the re-randomizer thread is $0.4\%$ for a period of 20~ms (shared across all modules).
A typical server has an average utilization $\approx$ 20-30\%~\cite{33387,bohrer2002case}. With the default Ubuntu configuration, a typical system uses around 100 modules.
Estimating very roughly, even if we make \textit{all} these modules re-randomizable and assume that CPU usage will increase by 0.36\% per every group of five additional modules as in ApacheBench (a very pessimistic, worst-case estimate), our approach can comfortably re-randomize over 950 modules.

\section{Security Analysis}
\label{sec:analysis}

\begin{description}[style=unboxed,leftmargin=0cm]

\item[Traditional ROP.]
Since attackers inject absolute addresses for ROP gadgets, and
the kernel uses one half of the 64-bit virtual address space
(57-bit for present CPUs), the probability of guessing a
correct address is $2^{-56}$, which is practically impossible.
Moreover, an attacker can often
assume that certain code is aligned at $2^{12}=4$KB page boundaries. Thus,
the effective probability is $2^{-(56 - 12)} = 2^{-44}$, which is still
unrealistic.
				In comparison, Shuffler's~\cite{SHUFFLER} and CodeArmor's~\cite{CODEARMOR} probability is only $2^{-(31 - 12)} = 2^{-19}$ (they use 32-bit offsets).

\item[Blind ROP] \cite{BLINDROP} uses \textit{fork(2)} and does not apply to the kernel.

\item[JIT ROP.]
The first step for the attack is to find some vulnerability such that ROP gadgets can be placed. We observe that: (1) vulnerabilities through buffer overflows are likely to be discovered in drivers, which we re-randomize; (2) stack re-randomization
and address encryption (Section~\ref{ref:stackreassign}) already prevent ROP gadget injections through the return address from non-rerandomized (e.g., kernel) code; (3) the corresponding module is re-randomized, i.e., code locations keep moving; (4) the latter implies that pointers are also adjusted when re-randomizing by adding an offset.\footnote{Those will typically be \emph{stack} and \emph{static data} pointers, e.g., the static ext4\_file\_inode\_operations structure (below) references many functions. \emph{Heap} pointers
are less common and are
modified during re-randomization since they are typically
module-local. See more details in the next paragraph.} If a hijacker inserts an absolute address to a non-rerandomized (e.g., kernel) code, this address is also adjusted and consequently becomes invalid due to completely random module movements. The only recourse for an attacker is to use non-rerandomizable (e.g., kernel) pointers inside the module. 
However, these pointers are strictly controlled by thin/secure wrappers (immovable part) and fixed GOTs (movable part). Fixed GOTs are write-protected, making hijacking impossible. Fixed GOTs are always accessed in RIP-relative mode, and the (movable-part) fixed GOT moves with the module (i.e., cannot be faked).
Moreover, JIT ROP attacks must complete between re-randomization 
intervals of modules, and an interval of one kernel module does not
coincide with that of another one, further reducing the probability
of inserting (and even discovering) a working chain of ROP gadgets.

The entire attack must be performed
within several milliseconds; all known attacks need several seconds to complete~\cite{SHUFFLER}.

\item[Address Hijacking.]
Aside from return-address hijacking, our approach defends against any \textit{static data} or \textit{stack} related attacks. In Linux, static data function pointers are very common; e.g., consider this definition in \path{fs/ext4/file.c}:
\begin{verbatim}
struct inode_operations ext4_file_inode_ops = {
    .setattr = ext4_setattr, ...        };
\end{verbatim}
Our idea stems from the fact that pointers that can be hijacked in a (re-randomizable) module can only be of two types: (1) pointers to addresses within the module itself (the vast majority of the pointers in almost all modules). This is not a concern; hijacking these pointers is futile since these addresses are going be changed randomly when the module is moved (re-randomized). (2) pointers to non-rerandomizable kernel code, e.g., network API calls, VFS calls, printk, etc.
The latter are guaranteed to be safely accessed through (read-only) GOT inside the
movable part since there is no other way to access arbitrary 64-bit addresses.

\begin{figure}[ht!]
\centering
\includegraphics[width=\columnwidth]{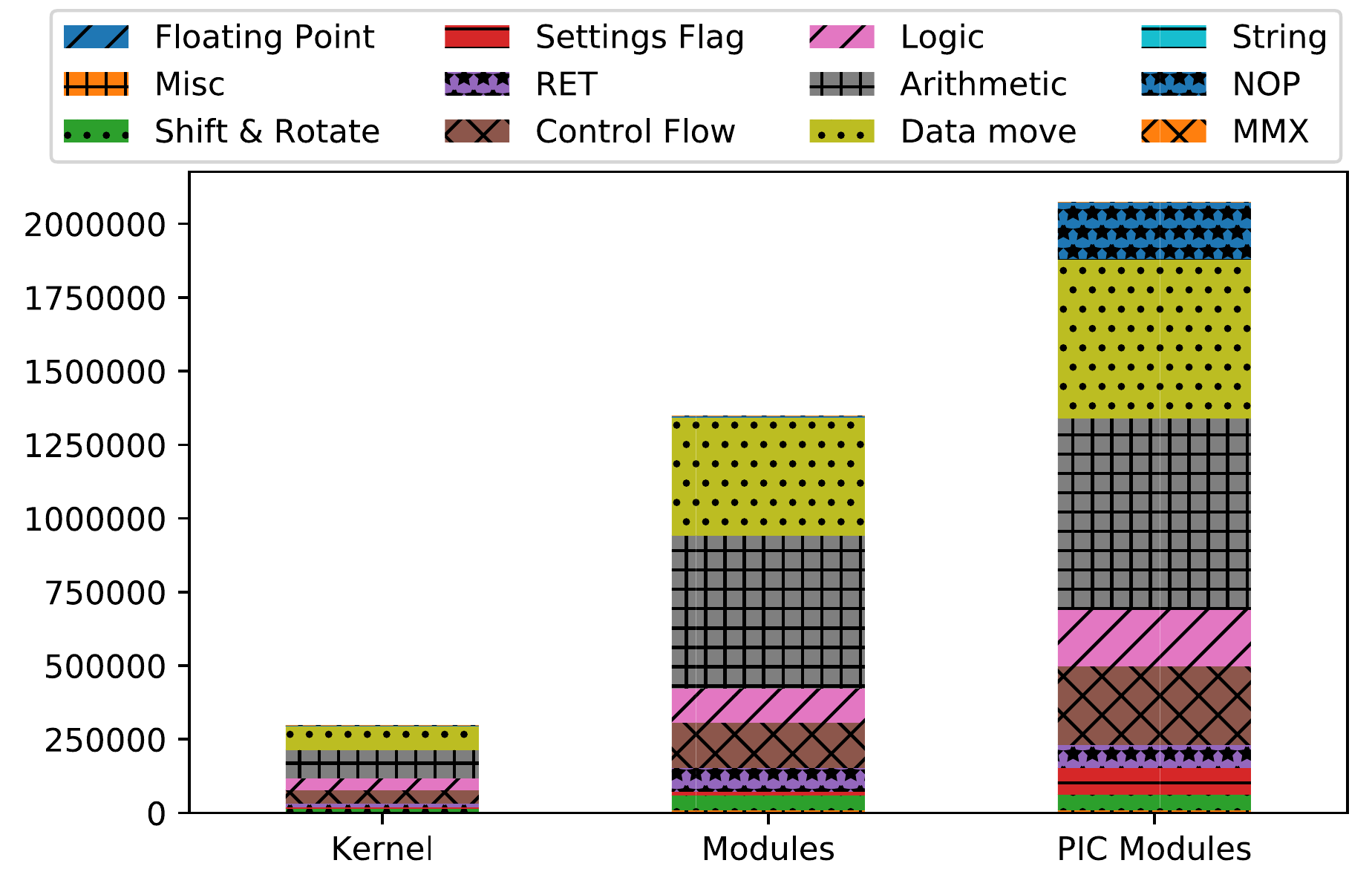}
\caption{ROP gadget distribution (the number of gadgets).}
\label{fig:gadget}
\end{figure}

The immovable part may also contain static data that exports \textit{wrapped} module functions. The wrappers will access module-local
functions through local GOT inside
the immovable part, which is read-only. Typically, suitable static data
itself is also read-only as in the example above (i.e., uses the \textit{const} qualifier). Even if the data is
writable, an attacker cannot easily modify
it since the immovable part
(as any other fixed kernel location) is arbitrarily far away from
the \textit{vulnerable} movable part.

A notable exception would be return addresses,
which are writable, not adjusted during re-randomization (unlike other module-local addresses), and
hard to manage in a controlled manner. For this reason,
we encrypt return addresses, making non-rerandomizable address
injections via them impossible.

Finally, even though \textit{heap-based} hijacking seems to be less of a
concern for the kernel in general (unlike typical user-space C++ code), it is still typically prevented by our approach. If a module allocates pointers in the heap (e.g., some data structure that is passed to the kernel), those addresses will be modified during re-randomization since they are typically module-local.
Commonly, modules \textit{export} their addresses to the kernel and \textit{import} only a few addresses (e.g., kernel API functions).

\item[Delayed Unmapping.]

A hypothetical problem arises when unmapping is delayed indefinitely. A similar concern exists in prior solutions, e.g., CodeArmor's QSBR, since unmapping cannot be triggered until pending calls are active.
We have not found that to be a practical issue for modules we tested, since all in-kernel calls are very quick and do not have indefinite blocking.
Linux also has built-in mechanisms to detect calls that block for too long, which safeguards from any potential issues.
In Section~\ref{sec:design}, we discuss how we address specific
corner cases such as long-term blocking and work queues so that
unmapping is never delayed indefinitely.

\item[ROP Gadget Distribution.]

We quantified ROP gadgets using the Ropper tool~\cite{schirra_2019}.
Our protection guarantees are not based on mere ROP gadgets availability, but their reduction
certainly makes the life of an attacker even harder.
Figure~\ref{fig:gadget} shows distribution for the kernel,
original modules, and our PIC-enabled modules in Ubuntu 18.04. The ROP
gadgets are classified according to the type of their instructions.
The immovable part of PIC modules has a negligible amount of gadgets as
almost all code stays in the movable part.
Most gadgets reside in modules, and only a fraction (15\%) is in the core kernel. The number of gadgets does increase for PIC vs. non-PIC modules, but PIC also enables 64-bit KASLR, efficient continuous re-randomization, and reduces the possibility of absolute address disclosures, which is a good trade-off.

To evaluate the quality of ROP gadgets, we constructed a specific example
with NX. Table~\ref{tbl:gadget_quality} shows that 80\% of the modules contain enough gadgets for a chain to disable NX.

\end{description}

\begin{table}
\caption{ROP gadget categories.}
		\label{tbl:gadget_quality}
		\begin{center}
\resizebox{.95\columnwidth}{!}{%
	\begin{tabular}{lcc}
     & \textbf{Non-PIC} & \textbf{PIC} \\
    \midrule
    \textbf{With ROP Chain, no side-effect} & 4,320 & 4,358 \\
    \textbf{With ROP Chain, with side-effect} & 1 & 1 \\
    \textbf{Without ROP Chain} & 1,008 & 970 \\
    \hline
    \textbf{Number of Modules} & 5,329 & 5,329 \\
    \end{tabular}
}
    \end{center}
\end{table}

\section{Related Work}
Modern attacks use ROP, even when using SGX~\cite{SGX}.
ASLR has been widely researched from early user-space implementations~\cite{MEMORYEXPLOIT,PAXASLR} to OS-specific
approaches such as fine-grained ASR in MINIX~\cite{ASRMINIX}. Similarly, (32-bit) KASLR was
implemented in Linux~\cite{KASLRLINUX}. Adelie enables full 64-bit KASLR in Linux by transforming all modules to use PIC.

Continuous re\hyp{}randomization was previously proposed and used in various contexts.
Stabilizer~\cite{STABILIZER} uses re\hyp{}randomization to enhance performance evaluation in user space. Stabilizer, however, focuses on performance evaluation and lacks Adelie's security features, e.g., address encryption.
Fine-grained ASR~\cite{ASRMINIX} is an early implementation of
re\hyp{}randomization for OSs that rely on component-based designs such as MINIX.
Shuffler~\cite{SHUFFLER} is a user-space technique for continuous re\hyp{}randomization to protect
against blind and just-in-time ROP attacks, a similar problem that we address for kernel code.
CodeArmor~\cite{CODEARMOR} is another user-space technique which uses page remapping and QSBR for better efficiency.
Shuffler's and CodeArmor's goals are different from ours: they rely on binary transformation of user-space programs,
whereas our technique benefits from Linux code availability
as well as from the position-independent model used by our modules.

There also exist (user-space) techniques that rely on compiler support. Remix~\cite{REMIX}
extends the LLVM compiler to add extra nop-paddings, allowing runtime flexibility
for moving code inside functions. This way ROP gadget locations change. However, attackers
can still use function pointers to defeat this re\hyp{}randomization scheme.
TASR~\cite{TASR} is another system based on the assumption that re\hyp{}randomization in the
program should happen before input and output (between corresponding system calls).
TASR is only suitable for user-space programs. Although authors
report that TASR overheads are very low,~\cite{SHUFFLER} argues there are
additional 30-40\% overheads due to the use of the -Og compiler optimization
flag (vs. -O2 which is normally used). The -Og flag is intrinsic to TASR.
TASR also has limitations unacceptable for typical kernel-mode code, e.g.,
disallowing upcasting into function pointers, use of custom memory allocators,
sizeof() assumptions, etc.

It is also possible to do arbitrary changes to existing binaries by a special recompiler. Egalito~\cite{10.1145/3373376.3378470} modifies existing binaries such that they can change process layout entirely. More specifically, Egalito is demonstrated to work together with JIT-Shuffling, a continuous re-randomization technique which is similar to TASR and Shuffler.

Although Adelie does not need to restart kernel modules while re-randomizing
address space layout, a number
of techniques to do so were proposed in the past. One approach~\cite{DRIVERRECOVERY}, which is specific to device drivers, enables user programs
to still work correctly even when device drivers are restarted. This approach extends the kernel
to use special shadow drivers which replace failed drivers. This approach cannot be
applied directly to enhance security, as an attacker can still figure out ROP gadget locations for shadow drivers and insert them.

Other techniques aim to enhance OS security by other means. NICKLE~\cite{KERNELROOTKITS}
uses virtual machines to run kernel code in shadow regions. The kernel code can be transparently
executed in a virtual machine in runtime. Similar techniques are also used by HookSafe~\cite{HOOKSAFE}
and hvmHarvard~\cite{HVMHARVARD}.
Other techniques such as~\cite{KERNELCONTROLDATA} rely on
special compiler support to protect kernel control data.
Unfortunately, none of these techniques provide an exhaustive solution to the security
problems that modern OSs have to deal with.

Adelie also differs from existing approaches in its focus on large-scale, monolithic OS
kernels such as Linux, which use a low-level programming model and have many inter-connected
components that lack strict boundaries and isolation from each other.
Adelie also provides a comprehensive solution that can be adopted to a wide range of modules.

Control Flow Integrity (CFI) can prevent code reuse attacks by ensuring that the control flow of a program remains valid. In CFI, any indirect branch taken by an application must be in accordance with its control flow graph.
CFI mechanisms were also applied to kernels (KCoFI)~\cite{Criswell:2014:KCC:2650286.2650804}.
KCoFI is compiler-based and can complement Adelie.
Unfortunately, CFI can still be defeated by a careful selection of ROP gadgets~\cite{davi2014stitching,evans2015control}.

Specialized hardware techniques, such as Morpheus~\cite{MORPHEUS}, can mitigate control-flow attacks. They can also complement Adelie's KASLR defense mechanisms for stronger security.

Finally, some papers~\cite{10.1145/3064176.3064216} focus on the NX bit enforcement and code diversification. This is still very desirable for Adelie, irrespective of continuous re-randomization, and can complement Adelie's defense mechanisms.

\section{Conclusions}

We presented Adelie, which contributes to KASLR in several ways.
First, we extend KASLR to 64 bits using PIC, which substantially increases KASLR's entropy and makes traditional ROP attacks impractical for kernel space.
It is intended for the entire kernel ecosystem. We successfully tested Ubuntu 18.04's default configuration (with $\approx$ 5000 modules) across different machines in a self-hosting mode.

The paper's other contributions are continuous re-random\-ization of kernel
modules, return address encryption, and stack re-random\-ization. This is the first effort to use these techniques in kernel space. Kernel space poses additional challenges due to a low-level API involving system calls, interrupt handling, and hardware access in device drivers. Due to the tremendous engineering effort involved, along with legacy and low-level code in the core kernel, we did not consider re\hyp{}randomizing the entire kernel. Moreover, re-randomization
incurs additional overheads. For these practical reasons, we re\hyp{}randomize only the most vulnerable components. 
However, as we justify in Section~\ref{sec:analysis}, we gain strong protection against ROP gadget injection for the entire kernel ecosystem regardless
of whether gadgets (e.g., in the core kernel) still exist.

Driver VMs, specialized OSs that run device drivers, exemplify one particular
use case where Adelie additionally strengthens the security of a system which is
already built with security considerations in mind. In fact, Adelie currently re-randomizes drivers in an enterprise-level system, which heavily relies on guest VM isolation
to protect against external and internal malicious actors. Adelie, however, is
more general and can be used for any Linux-based system.

Adelie is designed with Linux's source availability in mind.
Unlike Shuffler and CodeArmor, both user-space solutions, Adelie avoids binary-level transformation.
Also unlike them, Adelie implements 64-bit KASLR while using a zero-copying method for moving data and code, and efficiently keeps track of and unmaps previously used address ranges. Although CodeArmor also
employs remapping, it lacks PIC support and consequently restricts ASLR to 32 bits.
Adelie is also unique in the way it creates and handles multiple GOT tables.
Adelie's GCC plugin greatly reduces the engineering effort by changing kernel modules automatically.

\section*{Availability}

Adelie's latest code is available at \url{https://github.com/adelie-kaslr}.

\section*{Acknowledgements}

We would like to thank the anonymous reviewers and our shepherd Baris Kasikci for their insightful comments and suggestions, which helped greatly improve this paper.

This research is based upon work supported by the Office of the Director of National Intelligence (ODNI), Intelligence Advanced Research Projects Activity (IARPA). The views and conclusions contained herein are those of the authors and should not be interpreted as necessarily representing the official policies or endorsements, either expressed or implied, of the ODNI, IARPA, or the U.S. Government. The U.S. Government is authorized to reproduce and distribute reprints for Governmental purposes notwithstanding any copyright annotation thereon.

This research is also based upon work supported by the U.S. Office of Naval Research (ONR) under grants N00014-18-1-2022 and N00014-19-1-2493 and U.S. Naval Surface Warfare Center Dahlgren Division/NAVSEA/NEEC under grant N00174-20-1-0009. 

Adelie was integrated into an enterprise-level software infrastructure called SAVIOR (Secure Applications in Virtual Instantiations of Roles) system, which was developed as part of the IARPA VirtUE (Virtuous User Environment) program~\cite{IARPA}. SAVIOR's source code repository is publicly available~\cite{SAVIOR}. Adelie provides the Amazon ENA driver re-randomization in SAVIOR.

\appendix
\section{Artifact Appendix}

\subsection{Abstract}

In this Appendix, we provide information about how to deploy Adelie using
pre-installed VMs and run the benchmarks.
We also describe the hardware and software requirements necessary to run the experiments and reproduce the results presented in Section~\ref{sec:eval}.

Note that the experiments from the paper were run on physical machines, but we adapted the artifact to use VMs to make deployment and testing easier. As a result, we had to use E1000 rather than E1000E as a NIC driver. Also, the NVMe driver runs on top of an emulated rather than a physical device. Our artifact consists of two VMs: (1) the server VM with the modified Linux kernel and modules; (2) the client VM (load generator), where we keep benchmark scripts.

Our work specifically targets x86-64. Consequently, the provided artifact can only be deployed on x86-64 systems.

\subsection{Artifact check-list (meta-information)}

{\small
\begin{compactitem}
\item {\textbf{Program:} Modified Linux kernel and modules (Adelie). mySQL, Sysbench, Apache benchmarks are also included.}
  \item {\textbf{Compilation:} Ubuntu 18.04, GCC 8.4.}
  \item {\textbf{Transformations:} GCC plugins can be used for re-randomization.}
  \item {\textbf{Binary:} Pre-compiled kernel, kernel modules, and other binaries are provided in the VM images. }
  \item {\textbf{Run-time environment:} Ubuntu / VirtualBox. }
  \item {\textbf{Hardware:} Ideally, hardware described in Section~\ref{sec:eval} or similar. }
  \item {\textbf{Run-time state:} All VMs must be configured appropriately and placed on the same network. A correct host name, interface name, and other parameters must be specified on the client VM side. See below for more details. }
  \item {\textbf{Execution:} Automated via provided benchmark scripts. }
  \item {\textbf{Metrics:} Throughput. }
  \item {\textbf{Output:} Results are output to \verb|/home/client/benchmark/results| on the client VM. Additionally, the status of re-randomization can be analyzed by running `dmesg' (see below).}
  \item {\textbf{Experiments:} mySQL, Apache, IOCTL, Sysbench, etc. }
  \item {\textbf{How much disk space required (approximately)?:} 200GB (can be much less if not using FileIO).}
  \item {\textbf{How much time is needed to prepare workflow (approximately)?:} 30 minutes.}
  \item {\textbf{How much time is needed to complete experiments (approximately)?:} 2 hours (if not running many iterations).}
  \item {\textbf{Publicly available?:} Yes.}
  \item {\textbf{Code licenses (if publicly available)?:} The Linux kernel code uses Linux's original (GPLv2-only with the syscall exception) license unless specified otherwise. Other components that are not directly related to the Linux kernel (evaluation scripts, GCC plugins, etc) may be licensed under either the 3-Clause BSD or the GPLv2 license.}
  \item {\textbf{Archived (provide DOI)?:} \href{https://doi.org/10.5281/zenodo.5831326}{10.5281/zenodo.5831326}}
\end{compactitem}
}

\subsection{Description}

\subsubsection{How to access}

The artifact is available at \url{https://doi.org/10.5281/zenodo.5831326}. 

The artifact contains source code, benchmark scripts, and pre-installed VM images that should be used with VirtualBox. You need to download the client (load generator) VM image (\verb|Client.zip|) and the server VM image (\verb|Adelie.zip|). Adelie's latest source code is also available at \url{https://github.com/adelie-kaslr}.

\subsubsection{Hardware dependencies}

We recommend to use a system used for the paper evaluation (Xeon Silver 4114 2.20GHz, 96GB of RAM) or a system close to that.

\subsubsection{Software dependencies}

We used VirtualBox 6.1.26 to test the provided VM images. It should also be possible to use more recent versions of VirtualBox. VirtualBox must be used for both the client VM and the server VM images. For simplicity, both the client VM and the server VM can be executed on the same machine.

The provided VM images need VirtualBox Extension Pack. In Ubuntu,
VirtualBox and VirtualBox Extension Pack can be installed as follows:
\begin{verbatim}
$ sudo apt-get install virtualbox
$ sudo apt-get install virtualbox-ext-pack
\end{verbatim}

\subsection{Installation}

The installation and compilation process is non-trivial and time-consuming. Although the system was evaluated on physical hardware (which uses E1000E, NVMe), we decided to provide virtual images with similar drivers (E1000, NVMe). Although the results may not necessarily be as precise as in the paper due to virtualization, we went with this approach to simplify deployment.

Adelie's version of the Linux kernel and all other dependencies are installed on the provided VM images. Since the compiled Linux source tree occupies around 20GB, we did not include it into the server VM image. If you wish to compile Adelie's version of the Linux kernel yourself, you can follow the instructions in \verb|/home/asplos22/source/README.md| in the server VM image. We recommend using \verb|/home/asplos22/mnt| for compilation since the root file system does not have sufficient space. Source code is also provided outside of the VM image in \verb|source_code.zip|.

Before importing images into VirtualBox, you need to ensure that both images will use the same NAT network. First, go to \verb|File->Preferences->Network| in VirtualBox. Then, click on the plus sign. Create a new network named ``NatNetwork'' with CIDR ``10.0.2.0/24'' and enable DHCP.

Then, import both the server and client VM images into VirtualBox. The easiest way to import VMs is to extract files directly into the \verb|VirtualBox VMs| directory and then double click on the corresponding \verb|.vbox| files. Once imported, VM settings can be adjusted. We recommend increasing the default memory size, especially if you are planning to run the mySQL benchmark. At least 8GB is preferable. You may also adjust the number of virtual CPUs (vCPUs) accordingly. (The default parameters are 1GB of RAM and 1 vCPUs for the client VM, 4GB of RAM and 2 vCPUs for the server VM.)

Note that we already provide sample configuration files along with the VDI images. Please make sure that your configuration uses an E1000-compatible adapter (e.g., Intel 82540EM). The server VM also requires attention when importing storage images. The server VM must contain three storage images: \verb|Adelie_Boot| (the very first SATA image, the \verb|/boot| directory), \verb|Adelie| (the \verb|/| NVMe partition, which is re-randomized), \verb|Adelie_140G| (an additional 140GB-max NVMe partition which is used by FileIO tests and is mounted to \verb|/home/asplos22/mnt|).

When done, you can launch both the client VM and the server VM and log in.
\textbf{Username} and \textbf{password} for the server VM image are \verb|asplos22| and \verb|asplos22| correspondingly.
\textbf{Username} and \textbf{password} for the client VM image are \verb|client| and \verb|client|.

\subsection{Experiment workflow}

Running the experiments (and reboots) are automated through a script in the client VM. Please modify the configuration in the client VM (in the home directory):

\begin{verbatim}
$ vim benchmark/config.py
\end{verbatim}

Specifically, change $HOSTNAME$ to the IP address of the server (which can be identified by running `ifconfig' on the server side). Also specify the correct interface name in $ETH\_NAME$ (which can also be observed in `ifconfig'). You may also adjust other parameters via the config file, including which benchmarks you want to run (mySQL, Apache, etc.)

To run the benchmarks, use:
\begin{verbatim}
$ cd benchmark
$ python benchmark.py
\end{verbatim}

Once the benchmarks complete, you can optionally generate plots by running:
\begin{verbatim}
$ python plots.py
\end{verbatim}

\subsection{Evaluation and expected results}

Results are output to \verb|/home/client/benchmark/results| on the client VM. Additionally, if plots are generated, they will be located at \verb|/home/client/benchmark/plots|.

The benchmark script on the client VM will automatically reboot the server VM to load the correct version of the kernel. In the paper, we had different Linux kernels (e.g., retpoline enabled, retpoline disabled, etc). The differences are not very significant, thus we only provided one version of Adelie's 5.0.4-KASLR kernel with enabled retpoline and stack randomization. In this version, re-randomization can be enabled or disabled for the modules which were compiled with that option (e.g., e1000, nvme). Regardless of this, \textbf{all} modules use the position-independent code model as discussed in the paper. To further reduce the size of the server VM image, we also kept Ubuntu's out-of-the-box Linux kernel as the vanilla kernel.

Two different kernel versions will be evaluated by the benchmark scripts (vanilla and Adelie's KASLR). The 5.0.4-KASLR version will also run tests using re-randomization periods of varying lengths (i.e., 1 ms, 5 ms). 

Whenever the 5.0.4-KASLR kernel is loaded and the re-random\-ization period is non-zero, you should be able to see something like this by running `dmesg':
\begin{verbatim}
[  288.737300] Randomize: kthread started
[  289.023680] -----
[  289.023682] Randomized 53 times
[  289.023682] SMR Retire: 106
[  289.023683] SMR Free: 106
[  289.023683] SMR Delta: 0
[  289.023684] Stack Alloc: 530
[  289.023684] Stack Free: 530
[  289.023685] Stack Delta: 0
[  295.023676] -----
[  295.023677] Randomized 2117 times
[  295.023678] SMR Retire: 4234
[  295.023678] SMR Free: 4234
[  295.023679] SMR Delta: 0
[  295.023680] Stack Alloc: 21170
[  295.023680] Stack Free: 21170
[  295.023681] Stack Delta: 0
\end{verbatim}

Generally speaking, results corresponding to figures in Section~\ref{sec:eval} (mySQL, Apache, IOCTL, etc) are expected. The only excluded test is kernbench. It is not fundamentally important, but it would require storing (and compiling) the entire Linux source tree ($\approx$ 20GB) in the server image.

The benchmark scripts take care of re-randomization automatically. However, if desired, re-randomization of modules can also be started independently of the benchmark script (assuming that you boot up the 5.0.4-KASLR kernel). For example, to load the E1000 and NVMe modules with a re-randomization period of 20 ms, type:

\begin{verbatim}
$ sudo modprobe randmod \
>     module_names=e1000,nvme rand_period=20
\end{verbatim}

Similarly, to stop re-randomization, run:
\begin{verbatim}
$ sudo rmmod randmod
\end{verbatim}

\subsection{Experiment customization}

We recommend customizing tests in config.py. For example, if you only wish to run Apache, you can simply specify:
\begin{verbatim}
TESTS_TO_RUN = ["APACHE"]
\end{verbatim}

We also recommend reducing the number of iterations for Apache, especially if you need quicker testing. For this, you need to locate \verb|def run_apache_sweep| in \verb|benchmark.py| and change NUM\_REQ to 5000 (or use another smaller value).

\subsection{Notes}

We recommend adjusting the number of vCPUs and increasing RAM size for VMs. Our default configuration for VMs does not use too many cores and memory to be on the safe side for everyone. Please be advised that the output might not exactly match the paper results due to (1) using a machine with different specs than in the paper; (2) running all tests using VMs rather than natively.

If you encounter any errors during tests, try increasing the resources of both the server and client VMs. 

If desired, the contents of the VM images can be copied to physical partitions. Adelie can also be deployed to existing Linux installations by compiling from sources (see \verb|source_code.zip| and the provided \verb|README| file).

\subsection{Methodology}

Submission, reviewing and badging methodology:

\begin{sloppypar}
\begin{compactitem}
  \item \url{https://www.acm.org/publications/policies/artifact-review-badging}
  \item \url{http://cTuning.org/ae/submission-20201122.html}
  \item \url{http://cTuning.org/ae/reviewing-20201122.html}
\end{compactitem}
\end{sloppypar}

\balance

\bibliography{security}

\end{document}